\documentstyle[graphics]{mn}
\newcommand{\etal}{{\it et al.}~}

\newcommand{\eps}{\epsilon}

\newcommand{\bfx}{{\bf x}}
\newcommand{\bfr}{{\bf r}}

\newcommand{\bfv}{{\bf v}}

\newcommand{\bc}{\begin{center}}
\newcommand{\be}{\begin{equation}}
\newcommand{\ee}{\end{equation}}
\newcommand{\ec}{\end{center}}

\newcommand{\E}[1]{{\times 10^{#1}}}
\newcommand{\Mpc}{\rm Mpc}
\newcommand{\grad}{\nabla}
\newcommand{\apm}{APMSPH }
\newcommand{\hydra}{HYDRA }
\newcommand{\hydrastar}{HYDRA*}
\newcommand{\tree}{TREESPH }
\newcommand{\treestar}{TREESPH* }
\newcommand{\lya}{Lyman-$\alpha$~}
\newcommand{\eg}{{\it e.g.~}}
\newcommand{\ie}{{\it i.e.~}}
\renewcommand{\H}{{\mbox{${\rm H{\sc i}~}$}}}
\newcommand{\Hp}{{\mbox{${\rm H{\sc ii}~}$}}}
\newcommand{\He}{{\mbox{${\rm He{\sc i}~}$}}}
\newcommand{\Hep}{{\mbox{${\rm He{\sc ii}~}$}}}
\newcommand{\Hepp}{{\mbox{${\rm He{\sc iii}~}$}}}
\newcommand{\h} {{\rm H{\sc i}}}
\newcommand{\hp} {{\rm H{\sc ii}}}
\newcommand{\he} {{\rm He{\sc i}}}
\newcommand{\hep} {{\rm He{\sc ii}}}
\newcommand{\hepp} {{\rm He{\sc iii}}}
\newcommand{\ltsima}{\mbox{$\; \buildrel < \over \sim \;$}}
\def \simlt{\lower.5ex\hbox{\ltsima}}            
\def \gtsima{\mbox{$\; \buildrel > \over \sim \;$}}
\def \simgt{\lower.5ex\hbox{\gtsima}}            
\newcommand{\bbbb}{{\!\!\!\!}}
\title[P3M-SPH simulations of the \lya Forest]
{P$^3$M-SPH simulations of the \lya Forest}

\author[T. Theuns \etal]
{Tom Theuns$^{\,1}$, Anthony Leonard$^{\,2}$, George Efstathiou$^{\,1}$,
F. R. Pearce$^{\,3}$ and \newauthor P. A. Thomas$^{\,4}$\\
$^{\,1}$ Institute of Astronomy, Madingley Road, Cambridge CB3 0HA, UK\\
$^{\,2}$ Department of Physics, Astrophysics, University of Oxford, Keble Road, Oxford OX1 3RH, UK\\
${\,3}$ Department of Physics, University of Durham, South Road, Durham
DH1 3LE, UK\\ 
$^{\,4}$ Astronomy Centre, University of Sussex, Falmer, Brighton BN1
9QJ, UK}

\begin{document}

\maketitle

\begin{abstract}
We investigate the importance of several numerical artifacts such as
lack of resolution on spectral properties of the \lya-forest as
computed from cosmological hydrodynamic simulations in a standard cold
dark matter universe. We use a new simulation code which is based on a
combination of a hierarchical particle-particle--particle-mesh (P3M)
scheme for gravity and smoothed particle hydrodynamics (SPH) for gas
dynamics. We have performed extensive comparisons between this new code
and a modified version of the \hydra code of Couchman \etal and find
excellent agreement. We have also rerun the \tree simulations of
Hernquist \etal using our new codes and find very good agreement with
their published results. This shows that results from hydro dynamical
simulations that include cooling are reproducible with different
numerical algorithms. We then use our new code to investigate several
numerical effects such as resolution on spectral statistics deduced
from Voigt profile fitting of lines by running simulations with gas
particle masses of $1.4\times 10^{8}$, $1.8\times 10^{7}$, $2.2\times
10^{6}$ and $2.1\times 10^{5} M_\odot$. When we increase the numerical
resolution the mean effective hydrogen optical depth converges and so
does the column density distribution. However, higher resolution
simulations produce narrower lines and consequently the $b$-parameter
(velocity width) distribution has only marginally converged in our
highest resolution run. Obtaining numerical convergence for the mean
\Hep transmission is demanding. When progressively smaller halos are
resolved at better resolution, a larger fraction of low density gas
contracts to moderate over densities in which \Hep is already optically
thick, and this increases the net transmission, making it difficult to
simulate \Hep reliably. Our highest resolution simulation gives a mean
effective optical depth in \Hep 5\% lower than the simulation with
eight times lower mass resolution, illustrating the degree to which the
\Hep optical depth has converged. In contrast, the hydrogen mean
optical depth for these runs is identical. Since many properties of the
simulated \lya-forest depend on resolution, one should be careful when
deducing physical parameters from a comparison of the simulated forest
with the observed one. We compare predictions from our highest
resolution simulation in a cold dark matter universe with a
photo-ionising background inferred from quasars as computed by Haardt
\& Madau (1996), with observations. The simulation reproduces both the
\H column density and $b$-parameter distribution when we assume a high
baryon density, $\Omega_B h^2 \gtsima 0.028 $. In addition we need to
impose a higher IGM temperature than predicted within our basic set of
assumptions. We argue that such a higher temperature could be due to
differences between the assumed and true reionization history. The
simulated \H optical depth is in good agreement with observations but
the \Hep optical depth is lower than observed. Fitting the \Hep optical
depth requires a larger jump $\sim 14$ between the photon flux at the
\H and \Hep edge than is present in the Haardt \& Madau spectrum.
\end{abstract}

\begin{keywords}
cosmology: theory -- hydrodynamics -- large-scale structure of universe
-- quasars: absorption lines
\end{keywords}

\section{Introduction}
Sight lines to distant quasars intersect many cosmological structures
containing neutral hydrogen and \lya scattering by the \H in these
structures produces a forest of lines blueward of the quasar's \lya
emission line (Lynds 1971). This \lq\lya forest\rq~ contains unbiased
information on the temperature, density, velocity and ionization
structure of the intergalactic medium (IGM) along the line of sight to
the quasar, making the structures responsible for the \lya forest a
useful probe for studying the high-redshift universe. In addition, it
is likely that the absorbing gas retains a memory of its state at even
higher redshifts, enabling us to study its initial conditions (Croft
\etal 1998) and previous history. Since these structures are of
moderate density contrast, they are easier to simulate numerically than
galaxies, and consequently the high redshift universe can be studied
efficiently and accurately by comparing simulations of the \lya forest
with observations.

Recent hydrodynamic simulations of hierarchical structure formation in
a universe dominated by cold dark matter (CDM) have been shown to be
remarkably successful in reproducing a variety of statistics of \lya
absorption lines (Cen \etal 1994, Zhang, Anninos \& Norman 1995,
Miralda-Escud\'e \etal 1996, Hernquist \etal 1996, Wadsley \& Bond
1996, Zhang \etal 1997), including the number of lines per unit
redshift per unit column density and the number of lines with given
width (\lq $b$\rq~ parameter), as well as its evolution at low-redshift
(Theuns, Leonard \& Efstathiou 1998). This is quite encouraging for the
hierarchical picture of structure formation since the underlying
cosmological models were designed with galaxy formation in mind, hence
their \lya properties can be considered to be a genuine and successful
prediction. Most simulations to date have assumed a critical density,
cold dark matter model, in which a photo-ionising background close to
that inferred from quasars as computed by Haardt \& Madau (1996) is
required to explain the properties of the \lya forest. However, other
variants of the CDM model still provide acceptable fits, with only minor
modifications to the required photo-ionization background (Cen \etal
1994, Miralda-Escud\'e \etal 1996).

In this paper we introduce a new simulation code designed to study
numerically the formation of \lya systems. It is based on a combination
of Smoothed Particle Hydrodynamics (SPH, Lucy 1977, Gingold \& Monaghan
1977, see \eg Monaghan 1992 for a review) and an adaptive P3M
(particle-particle--particle-mesh) gravity solver (Couchman 1991). Its
efficient gravity solver and SPH implementation lead to a fast and
accurate code which has the potential to extend considerably the
dynamic range of the simulations. We discuss tests of the new code and
perform extensive comparisons against two other simulation codes:
\hydra and \tree. Both of these are also based on SPH but their gravity
solvers differ: \hydra \cite{Couchman95} uses the same gravity solver
as \apm but \tree \cite{KatzWeinbergHernquist96} uses a tree
structure. We discuss in detail the differences between the \apm and
\hydra codes. We also discuss the changes we have made to the
publically available \hydra code to study the \lya cloud problem. The
overall agreement between the three codes is excellent which shows that
hydrodynamic simulations that include cooling are reproducible with
different simulation codes. The good agreement also shows that \hydra
can be used to study the \lya problem and we are currently analysing
several large \hydra simulations performed on the T3D computer to
understand in more detail how resolution affects \lya statistics (the
VIRGO consortium, in preparation).

We then use \apm to perform simulations at increased resolution and
establish the extent to which published results are influenced by lack
of numerical resolution and other numerical artifacts. Wadsley \& Bond
(1996; see also Bond \& Wadsley 1997) recently warned simulators of the
importance of long-wavelength perturbations on the occurrence of
filamentary structures in simulations. This is illustrated explicitly
in the work of Miralda-Escud\'e \etal (1996) who compare simulations
with the same resolution but different box sizes. Unfortunately,
current numerical codes do not possess the required dynamic range to
resolve the Jeans length in a very large simulation box. We try to
gauge the effects of missing waves and of failing to resolve the Jeans
length by performing simulations with various box sizes. This paper is
organised as follows: Section~\ref{sect:simulation} discusses the
physical model and gives details of the simulation codes,
Section~\ref{sect:comparison} presents the comparisons between codes,
Section~\ref{sect:resolution} addresses the importance of numerical
resolution, Section~\ref{sect:observations} does a comparison of
simulations against observations and finally Section~\ref{sect:summary}
summarises. Technical details are relegated to Appendices.

\section{Simulation}
\label{sect:simulation}
\subsection{Physical model}
\label{sect:physmodel}
We model the evolution of a periodic, cubical region of a critical
density Einstein-de Sitter universe ($\Omega=1$,
$\Omega_\Lambda=0$). The Newtonian equations of motion governing the
evolution of structures are given in Appendix~A. The comoving size of
the simulated box is $L/(2h)$ \Mpc, where the Hubble constant today is
written as $H_0=100 h$ km s$^{-1}$ Mpc$^{-1}$. We will assume $h=0.5$
throughout and describe simulations with $L=2.5$, $L=5.5$, $11.11$ and
$22.22$\Mpc. A fraction $\Omega_B=0.05$ of the matter density is
assumed to be baryonic, consistent with limits from nucleosynthesis
(Walker \etal 1991, but note the continuing debate on the deuterium
abundance derived from quasar spectra, favouring higher values
$\Omega_B\approx 0.075$, see \eg Burles \& Tytler 1997 and references
therein). The rest of the matter is in the form of cold dark matter
(DM). These model parameters were chosen to enable comparison with
the \tree simulation of Hernquist \etal (1996) which has identical
parameters to our lowest resolution run. We use the smaller boxes which
have correspondingly higher resolution to test for numerical
convergence. The simulations are started at a redshift $z=49$ and we
follow the evolution to $z=2$. To generate initial conditions for the
simulations we use the following fit to the CDM linear transfer
function from Bardeen \etal (1986)
\begin{eqnarray}
T(k) &=&(1+3.89q+(16.1q)^2+(5.46q)^3+(6.71q)^4)^{-1/4}\nonumber\\
&\times& {{\rm ln} (1+2.34q)\over 2.34 q}\,,
\end{eqnarray}
where $q= k/h^2{\rm Mpc}$ and normalise it such that $\sigma_{8}=0.7$.
Here, $\sigma_8^2$ denotes the r.m.s. of mass fluctuation in spheres of
radius 8$h^{-1}$~Mpc. This value of $\sigma_8$ is higher than the one
deduced from the abundance of galaxy clusters ($\sigma_8=0.52\pm 0.04$,
Eke, Cole \& Frenk 1996) but we use it to allow a direct comparison
with the Hernquist \etal (1996) simulations. The equations of motion
used in the SPH code are detailed in Appendix~\ref{sect:eqs}.

Gas is allowed to interact with the cosmic microwave background
radiation (CMB) through Compton cooling and with an imposed uniform
background of ionising photons, assumed to originate from quasars
and/or young galaxies. Collisions between atoms that lead to ionization
represent a loss term for the optically thin gas causing cooling,
whereas photo-ionizations heat the gas because the ionised electron
carries excess kinetic energy. We detail the temperature dependence of
the cross-sections for these processes in Appendix~\ref{sect:cooling}.
They are taken from Cen (1992) with some minor modifications. The flux
spectrum of the ionising photons from a given quasar source seen by an
average \lya cloud is changed due to reprocessing (absorption and
emission) by intervening clouds in the clumpy IGM. The amplitude of the
ionising background changes due to the evolution of the quasar
luminosity function, causing the flux spectrum to depend on
redshift. Haardt \& Madau (1996) took all these effects into account
and provide fits to the photo-ionization and photo-heating rate as a
function of redshift for the case where the main sources of UV photons
are quasars. We use the Haardt \& Madau rates but divide them by two so
that they are identical to the rates used in the \tree simulations of
Hernquist \etal (1996), enabling us to compare directly with their
results. The imposed background flux is time-dependent but we assume
nevertheless that the gas remains in ionization equilibrium throughout
(see below). The resulting fits to the photo-heating and
photo-ionization rates can be found in Appendix~\ref{sect:cooling} and
we will refer to them as \lq HM/2\rq. We will also present simulations
with an ionising background of radiation with constant amplitude
$J_{21}$ and power law spectrum with index $\alpha$ (see
Appendix~~\ref{sect:cooling} for details). $J_{21}$ denotes the
amplitude of the radiation spectrum at the hydrogen \lya edge in the
usual units ($10^{-21}$ ergs cm$^{-2}$ s$^{-1}$ Hz$^{-1}$
sr$^{-1}$). The net cooling and heating rates depend on the relative
helium abundance by mass for which we assume $Y=0.24$.

\begin{figure}
\resizebox{\columnwidth}{!}{\includegraphics{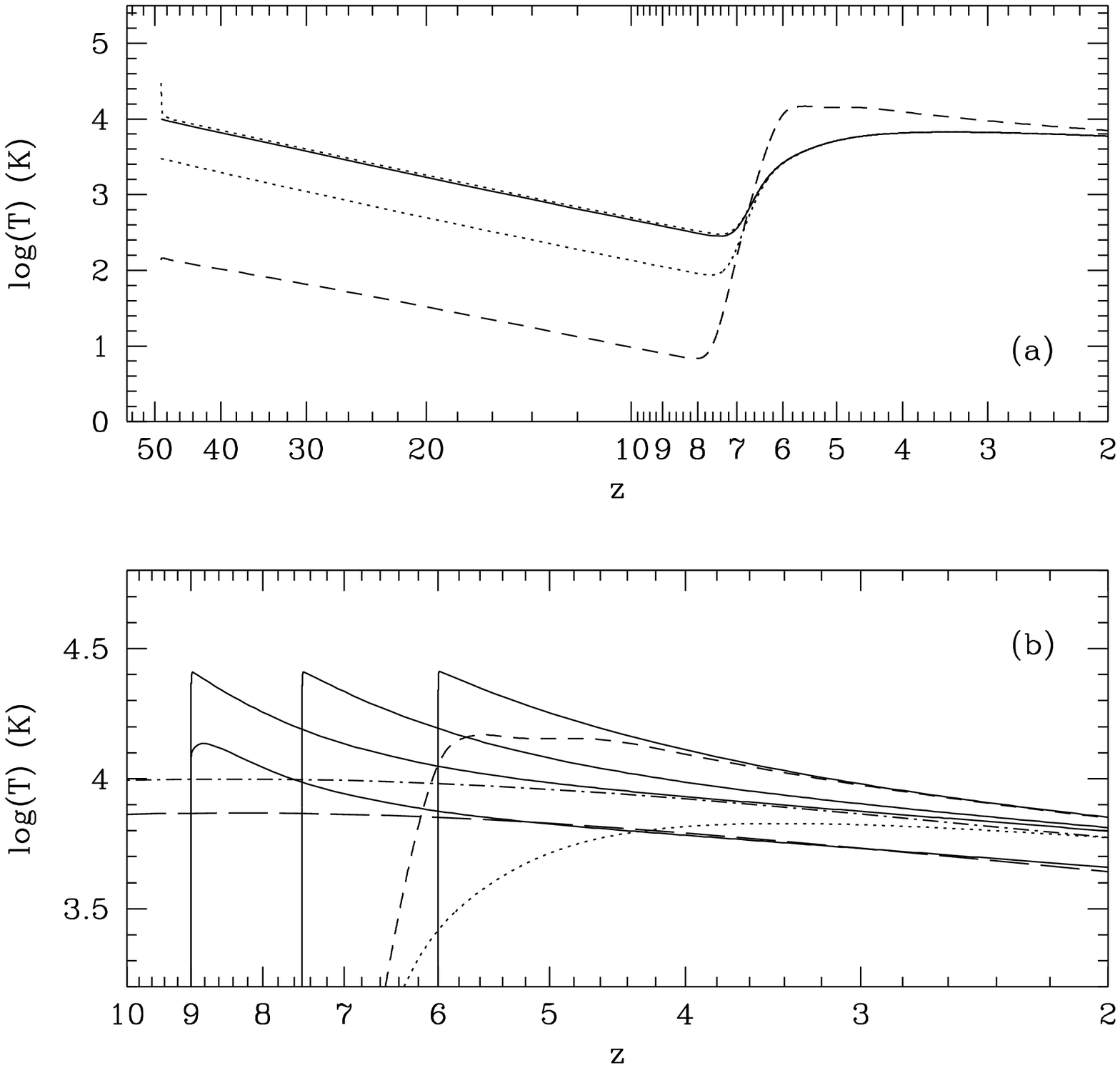}}
\caption{Evolutionary tracks of IGM temperature at $\rho/\bar{\rho}_B =
1$ ($\Omega=1, h=0.5, \Omega_B=0.05, Y=0.24$). (a) Evolution from
$z=49$ assuming the background ionising flux as computed by Haardt \&
Madau (1996) but with amplitude divided by 2, ionization equilibrium
and setting initial temperatures, $T_i = 10^4$ K (solid), $10^{4.5}$
and $10^{3.5}$ K (both dotted). The evolution in the non-equilibrium
case, with $T_i = 2.74(1+49)$ is also plotted (dashed). Both cases are
also shown in the lower panel for comparison (dotted and dashed
respectively). (b) Temperature evolution including non-equilibrium
effects resulting when a constant ($J_{21}= 0.5, \alpha =1$) power-law
UV background is `switched on' at $z_{on}$ = 9, 7.5 and 6. These
converge with decreasing $z$ to the temperature where the net cooling
time equals the Hubble time (dot-dashed). We also show evolution with
$z_{on}=9$ assuming $\alpha = 3$, similarly converging on the (long
dashed) temperature for which the net cooling time equals the Hubble
time for that radiation spectrum.}
\label{fig:tempigm}
\end{figure}

Since our simulations include gas we need to specify the initial
temperature of the IGM at high redshift. We will first describe the
temperature evolution in the absence of any extra energy input and then
discuss various reionization scenarios. These models are shown in
figure~\ref{fig:tempigm} and were computed using a scheme based on
Giroux \& Shapiro (1996). At high redshifts, Compton cooling is very
efficient hence gas will cool very quickly due to the coupling of the
free electrons left over from recombination to the CMB photons giving
$T\sim T_{CMB}=2.7\times (1+z)$K. At later times, the gas becomes
progressively more neutral to make the coupling with the CMB
inefficient, causing the temperature to drop almost adiabatically,
$T\propto (1+z)^\alpha$, with $\alpha\sim 1.8$ (Giroux \& Shapiro
1996). This temperature evolution is shown as the dashed line in
figure~\ref{fig:tempigm}. In contrast, in the simulations described
here, we assume an IGM temperature at the mean IGM density of $T=10^4$K
at $z=49$ and in addition assume ionization equilibrium, in which case
the gas will cool adiabatically $T\propto (1+z)^2$ (full line in the
upper panel). If the starting temperature is higher, the gas will cool
very efficiently through Compton cooling until it becomes mostly
neutral at $T\sim 10^4$K, after which it cools adiabatically (upper
dotted line). If the starting temperature is lower, it will cool
adiabatically from the start (lower dotted curve). In any of the
previous cases, $T$ will be low with respect to the reionization value
provided reionization occurs at redshifts $z\le 10$ say, hence $T$
after reionization is independent of our assumed starting temperature
at $z=49$.

The behaviour of $T$ {\em after} reionization depends on the
reionization scenario and to some extent on whether non-equilibrium
ionization effects are taken into account, as is illustrated in
figure~\ref{fig:tempigm} (lower panel). If reionization occurs
impulsively, more ionizations will take place per unit time than in
equilibrium conditions so the non-equilibrium gas will become hotter
(Miralda-Escud\'e \& Rees 1994, Haehnelt \& Steinmetz 1998). Since the
thermal time scales are long in low density gas, this difference in
temperature may survive to low redshifts. The IGM temperature is
sensitive to the imposed photo-ionization heating rate, notably the
\Hep heating rate at reionization, which is based on an uncertain
extrapolation to $z\sim 6$ of the quasar luminosity function. This
introduces a considerable uncertainty in the temperature of the IGM,
even at redshifts as low as $\sim 2$.

The temperature of the non-equilibrium gas will approach that of the
equilibrium gas if the flux of photo ionising photons evolves slowly
after reionization so that ionization equilibrium is
re-established. From then on, photo ionization heating competes with
Compton cooling and adiabatic expansion in trying to change the gas
temperature. The age of the universe at a given redshift then basically
determines how long the gas has been heated, which in turn determines
its temperature. This sequence of events is illustrated in
figure~\ref{fig:tempigm}b for a range of reionization scenarios. In the
low density regime, the gas temperature depends on its density in a
straightforward way which is derived in Appendix~\ref{App:IGM} (see
also Giroux \& Shapiro 1996 and Hui \& Gnedin 1997). The equilibrium
temperature of the IGM, \ie that temperature where heating balances
cooling, is generally higher than this (\eg Zhang \etal 1997). In
contrast, at high densities line cooling in the shock-heated gas
becomes important and the gas cools to temperatures of a few times
$10^4$K, below which atomic hydrogen cooling becomes inefficient. Since
we do not include molecular hydrogen, the gas in the simulations cannot
cool radiatively below $T=10^4$K. In summary (for $z\sim 6\rightarrow
2$, after reionization): for low densities ($\le
0.1\langle\rho_B\rangle$) there is a one-to-one relation between $T$
and $\rho$ obeyed by unshocked gas which is determined by the photo
ionization heating rate, adiabatic cooling, and to a small extent by
Compton cooling. At high densities, $\rho \ge 10^2\langle
\rho_B\rangle$, collisions try to cool the shock-heated gas to $T\sim
10^4$K. At intermediate densities, there is a large range in $T$ for
any given $\rho$, depending on the extent to which the gas has been
shocked.

\subsection{Simulation methods}
The numerical methods described here are based on Smoothed Particle
Hydrodynamics (SPH, Lucy 1977, Gingold \& Monaghan 1977, see \eg
Monaghan 1992 for a review) for hydrodynamics and P3M (Hockney \&
Eastwood 1988, Efstathiou \etal 1985) for self-gravity. The gas in the
simulation is represented by a set of SPH particles which each carry
the same mass but possibly a different thermal energy. Spline
interpolation over these particles allows one to compute smoothed
estimates for density and temperatures throughout the computational
volume. Gradients may also be computed. The width of the spline kernel
is matched to the local particle number density and so high density
regions have higher numerical resolution than do low density regions,
in contrast to Eulerian schemes. P3M uses a combination of Fast Fourier
Transforms (FFTs) and local direct summation to combine speed with
accuracy. We compare detailed results from two codes. \apm was written
specifically for this problem by one of us (TT) and is based on the
hierarchical P3M code of Couchman (1991). \hydra \cite{Couchman95} is a
publically available code used extensively by the VIRGO consortium. We
have modified \hydra to include photo-heating effects and to improve
the simulation of low density regions (Section~\ref{sect:hydra} and
Appendix~\ref{sect:densityfix}). We also compare with the published
results of the \tree code
\cite{HernquistKatz89,KatzWeinbergHernquist96}. In the rest of this
section we will describe some technical details of the codes pertinent
to their comparison.

\subsection{\apm}
\label{sect:apm}
This code is based on the adaptive P3M implementation of Couchman
(1991) which uses mesh refinements in regions of high particle number
density to speed-up gravity particle-particle interactions. The SPH
implementation uses a similar but separate linked-list scheme based on
a hierarchy of grids to find neighbouring SPH particles. This is
because the refinements used by the gravity part of the code are not
necessarily optimal for the SPH calculation and vice versa. Note that
in \apm we find the neighbours of all particles even in low density
regions, in contrast to other P3M+SPH implementations such as \hydra
(see below) and Evrard's (1988) versions. The explicit expressions used
to compute the SPH accelerations are given in Section~\ref{sect:SPH} as
well as details of our method of determining SPH neighbours.

Given a power spectrum we set-up initial conditions by perturbing
particles from a grid using the Zel'dovich (1970) approximation
(Efstathiou \etal 1985). We take the DM and SPH particle grids offset
by half a cell size. We then march the particle positions forwards in
time using a second-order accurate leap-frog integrator with variable
time-step, using the correction procedure of Hernquist \& Katz (1989)
to keep the scheme second-order even when the step changes.  The SPH
accelerations depend on the particles' velocities: we synchronise these
by predicting velocities over half a time-step. As is usual, we take a
time-step based on the Courant-Friedrichs-Levy condition (CFL, \eg
Monaghan 1992), but take smaller steps whenever violent shocks
occur. Such shocks are flagged by large values of the artificial
viscosity terms (\eg Katz \etal 1996b) and the latter decrease the
allowed time-step for accuracy and numerical stability (\eg Hockney \&
Eastwood 1988, ch. 4). Since we use a uniform time-step, we take as
system time-step the minimum time-step over all particles. To avoid one
or a few particles from unnecessarily slowing down all the others, we
make sure that there are a reasonable number of particles with
similarly small time-step, by increasing the resolution length of those
few offending particles that would otherwise require an even smaller
step by a factor $\ltsima 1.2$.

The integration of the cooling terms requires prohibitively small
time-steps, comparable to the local cooling time. We solve this problem
in the usual way by integrating the thermal energy equation at with an
implicit scheme, assuming in addition ionization equilibrium and fixed
density. We evaluate the cooling and heating rates by interpolation
from tables which we recompute every time the background flux changes
significantly.

This new SPH+AP3M implementation was tested with a variety of test
problems: the 1D shock tube, the spherically symmetric collapse of a
gas sphere and the formation of a massive galaxy cluster\footnote{We
are indebted to Adrian Jenkins for providing us with the initial
conditions and final profiles for this problem as computed using
\hydra}. Energy conservation based on the the Layzer-Irvine cosmic
energy (equation~\ref{eq:Layzer} below) gives typically $\Delta I/W\sim
1-2\%$ for the runs discussed here. A simulation with $64^3$ SPH and
equal number of DM particles takes $\sim$ 1000 steps to evolve from
$z=49$ to $z=2$ for a box size of 22Mpc. Using a $128^3$ grid for the
gravity calculations, the code takes 110$s$ per step at $z=50$ (40$s$
for gravity, $50$ for SPH) which increases to 230$s$ at $z=2$ (125$s$
for gravity, $82$ for SPH) on a DEC-alpha 4100 server. The total
simulation then takes $\approx$ 32 hours and requires $\approx$ 60
M bytes of RAM memory in single precision (32 bits).

\subsection{\hydra}
\label{sect:hydra}
\hydra (Couchman \etal 1995) is similar to \apm in that it combines a
variable resolution SPH code with the adaptive P3M code of Couchman
(1991). This code has been used extensively by the VIRGO Consortium on
massively parallel computers such as the Cray T3D \& T3E computers in
Edinburgh and Munich, with the primary aim of studying large scale
structure (Jenkins \etal 1997), cluster (Colberg \etal 1997) and galaxy
formation problems (Pearce \etal 1998), where typically very large
dynamic ranges are required. Throughout its development the code has
also been used to explore the accuracies and inaccuracies of the SPH
technique in modelling astrophysical scenarios, in keeping with the
spirit of this paper. In this section we concentrate on reporting the
code-modifications introduced to allow an accurate treatment of
photoionised primordial gas dynamics on scales of $\gtsima$ 10kpc.

The first modification involved updating the thermal energy solver to
take into account the cooling and heating functions in ionization
equilibrium, as listed in Appendix~\ref{sect:cooling} and also used in
\apm. We then tested this new solver by comparing with the \apm one and
another solver developed by one of us (APBL, based on a scheme of
Giroux \& Shapiro 1996), for a wide range of initial temperatures,
densities and photo ionising background choices.

Since the \lya forest absorption is expected to arise from gas in the
IGM at moderate over densities and slight under densities, it is
paramount that an accurate handling of low-density gas is achieved in
the simulations. \hydra suffers in this respect since its
implementation requires that the same linked-list to find neighbours be
used for both gravity and SPH interactions. This leads to the following
problem: if a linked-list is used that is optimal for the gravity
calculation, then particles in low density regions will not find enough
SPH neighbours to accurately compute hydrodynamic quantities such as
their density. If, on the other hand, a linked-list is used which is
suited to the SPH computation, then the gravity calculation will become
very time-consuming.  We decided to deal with this problem in a
two-step fashion. Firstly, we run our simulations with FFT cells for
the base mesh (the PP linking cells are typically 2-3 FFT cells on a
side) set at half the mean inter-particle separation (leading to an
efficient gravity calculation) but use a correction for the SPH
smoothing kernel applied for particles with few neighbours (see
below). Secondly, post-simulation, we continue every end state for two
further time-steps where the number of linking cells is decreased by a
factor $4^3$ and the original SPH kernel is used to re-calculate the
densities for all particles with $\rho/\bar{\rho}_B \ltsima 10$ from
their respective positions as simulated. Only particle densities are
ever updated in this step, referred to as the \lq density
reconstruction\rq~ step.

Since the temperatures in the low-density IGM are density dependent
these must be post-adjusted during density reconstruction as well.
Fortunately we can use the power-law density temperature relation
obeyed by low density gas (see figure~\ref{fig:rtclose} below). This
relation is history independent if reionization occurs gradually or at
sufficiently high redshift, as discussed in
Section~\ref{sect:physmodel}. In Section~\ref{sect:comparison} we shall
compare the effects of this scheme with the exact density scheme of
\apm, and show that the temperature-density distributions simulated by
the two codes are essentially indistinguishable, also at low
densities. More details of this \hydra low density correction are given
in Section~\ref{sect:densityfix}.
  
\subsection{\tree}
\tree \cite{KatzWeinbergHernquist96} uses a tree structure to compute
gravitational forces using multipole expansion based on an accuracy
criteria and to find SPH neighbours. Each SPH particle is advanced with
its own time-step based on the CFL condition, which may be smaller than
the \lq~system\rq~ time-step imposed on all the collisionless
particles. \tree uses the same spline kernel as \apm and \hydra but a
different symmetrization procedure to compute SPH pairwise quantities
(Katz \etal 1996b). Gravitational forces are softened using a
spline-kernel softening and made periodic using the Ewald summation
method. We will compare our results with the published \tree simulation
results of Hernquist \etal (1996).

\section{Code comparisons}
\label{sect:comparison}
The characteristics of the absorption lines in simulated spectra
reflect the properties of the neutral gas in the simulation box. For
example, stronger lines tend to be produced in regions with higher
densities and correspondingly higher temperatures. For such lines,
effects of peculiar velocities and Hubble expansion are less important
and deviations from Voigt profiles are usually caused by surrounding
matter that often lies in a filamentary distribution and contributes to
the absorption.  Consequently, the statistical properties of strong
lines mostly reflect the statistics and shapes of halos that form in
the simulation. In contrast, weaker lines usually originate in
filaments and pancakes and residual Hubble expansion contributes
significantly in shaping the line. These lines then reflect the
properties of the gas distribution in the low density IGM. We will
first compare the overall gas distribution between \apm and \hydra by
studying the amount of cold gas responsible for most of the weaker
lines before concentrating on the halo statistics. In the second part
of the section, we will compute and compare simulated spectra. These
will be analysed using Voigt profile fitting in the same way as
observed spectra.

To make comparison with the \tree runs (Hernquist \etal 1996) we have
run simulations with initial conditions computed from the initial
linear density field originally used by these authors which they kindly
provided for us. We shall refer to these runs as 22-64-k, where 22
refers to the box size (22.22 comoving Mpc) and 64 to the cube root of
the number of particles (\ie $64^3$ particles of a single species) in
the simulation. Throughout this paper we will compare results from
simulations using a variety of different codes and numerical parameters
and refer to them as above. These are summarised in
Table~\ref{table:runs} and are all performed in a standard CDM
cosmology ($\Omega=1$, $h=0.5$). In what follows, \apm simulations will
be labelled with \lq A\rq~ and the \hydra ones with \lq H\rq.

\begin{table*}
 \centering \begin{minipage}{140mm} \caption{Runs performed. All runs
 assume a standard adiabatic, scale-invariant CDM cosmology
 ($\Omega=1$, $\Omega_\Lambda=0$, $H_0=50$ km s$^{-1}$ Mpc$^{-1}$) with
 normalisation $\sigma_{8}=0.7$ and baryon fraction
 $\Omega_B=0.05$. Each run had equal numbers of SPH and dark matter
 particles. HKWM refers to Hernquist \etal (1996) initial
 conditions. $\epsilon$ is the comoving softening for the spline
 gravity kernel, the SPH resolution length $h$ is not allowed to fall
 below $\epsilon/2$. \lq HM/2\rq~ refers to the radiation spectrum
 computed by Haardt \& Madau (1996), with amplitude divided by
 2. $J_{21}=0.5$, $\alpha=1$ refers to a power law spectrum of ionising
 photons. cen and apb refer to different fits to the cooling and
 heating functions (see Appendix \ref{sect:cooling}).}
 \begin{tabular}{@{}lllllllll@{}} & Name & Box size (Mpc) & $N_{\rm
 SPH}$ & ICs & code & $\epsilon$ (kpc) & $J(z)$ & rates \\[10pt] 1 &
 22-64-k & 22.22 & $64^3$ & HKWM & \apm,\hydra & 17.3,20 & HM/2 & cen\\
 2 & 22-64 & 22.22 & $64^3$ & own & \apm,\hydra & 17.3,20 & HM/2 &
 apb\\ 3 & 11-64 & 11.11 & $64^3$ & own & \apm,\hydra & 20,10 & HM/2 &
 apb\\ 4 & 5-64 & 5.5 & $64^3$ & own & \apm & 10 & HM/2 & apb\\ 5 &
 2.5-64 & 2.5 & $64^3$ & own & \apm & 4.5 & HM/2\footnote{For this run
 we imposed the $z=5$ HM/2 background for all higher $z$ as well. We
 stop this run at $z=4$.} & apb\\ 6 & 22-32 & 22.22 & $32^3$ & own &
 \apm,\hydra & 17,20 & HM/2 & apb\\ 7 & 11-32 & 11.11 & $32^3$ & own &
 \apm,\hydra & 8.6,10 & HM/2 & apb\\ 8 & 5-32 & 5.5 & $32^3$ & own &
 \apm & 4.3 & HM/2 & apb\\ 9 & 11-64-j & 11.11 & $64^3$ & own & \hydra
 & 10 & $J_{21}=0.5$, $\alpha=1$ & apb\\
\label{table:runs}
\end{tabular}
\end{minipage}
\end{table*}

\subsection{Gas distribution}
\subsubsection{Global distribution}
\label{sect:globalgas}
\begin{figure*}
\setlength{\unitlength}{1cm}
\centering
\begin{picture}(10,8)
\put(-1.,-6.0){\includegraphics{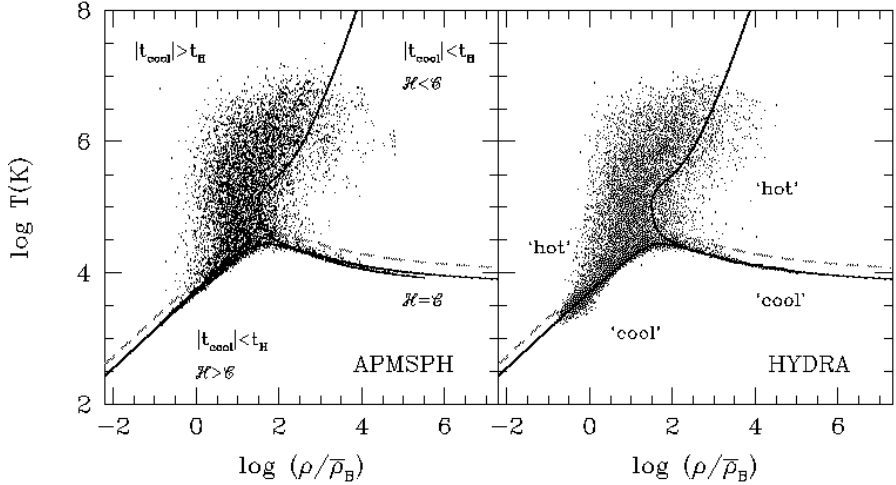}}
\end{picture}
\caption{Temperature-density distribution for runs 22-64-k at $z=2$ for
\apm (left panel) and \hydra (right panel).  The small differences in
the low density, low temperature gas distribution between \apm and
\hydra are a consequence of the low density correction applied to the
\hydra simulation, as discussed in the Appendix. The solid line
indicates the locus where the cooling time equals the Hubble time,
$t_{\rm cool}=t_{\rm H}$. At high densities, this corresponds to the
equilibrium temperature where cooling balances heating. The dashed line
shows 1.5$\times T_{\rm min}(\rho)$, where $T_{\rm min}$ is defined as
that temperature below which heating is dominant and on which the net
cooling time is equal to the Hubble time (see text for details). At
high densities, $T_{\rm min}$ equals the equilibrium temperature. Only
1 particle in 16 is plotted. }
\label{fig:figrt}
\end{figure*}

In figure~\ref{fig:figrt} we present the temperature-density
distribution for the simulations A-22-64-k and H-22-64-k at $z=2$. The
overall distributions look very similar and can be understood in terms
of the relative efficiencies of cooling and heating and the respective
time scales involved, as will be described below. We first define the
cooling time of gas at temperature $T$ as
\begin{equation}
t_{\rm cool}={u\over |\dot{u}|} = {3 k_B T\over 2 \mu}
{m_H \over \rho(1-Y)^2 ({\cal C}-{\cal H})}\,.
\label{eq:tc}
\end{equation}
where $k_B$ is Boltzmann's constant, $\mu$ is the mean molecular weight
and $m_H$ is the proton mass. ${\cal C}(u)$ and ${\cal H}(\rho, u)$ are
the normalised cooling and heating rates of
Appendix~\ref{sect:cooling}. The Hubble time is $t_{\rm H}=(6\pi
G\bar{\rho}(z))^{-1/2}$.  On the left hand panel of
figure~\ref{fig:figrt} we indicate with a solid line the location where
the absolute value of the cooling time equals the Hubble time at
$z=2$. This line splits the diagram into three separate regions. At low
densities and high temperatures the cooling time is longer than the
Hubble time, \ie neither heating nor cooling are able to change the gas
temperature significantly. At high densities and high temperatures,
bremsstrahlung and line cooling processes always dominate leading to
cooling times shorter than the Hubble time. Finally, at low densities
photo-heating is dominant leading to heating times shorter than the
Hubble time. (For ${\cal H}<{\cal C}$ the gas is heated and the cooling
time in equation~(\ref{eq:tc}) is negative. In such a case we call
$-t_{\rm cool}$ the heating time, and $|t_{{\rm cool}}|$ the \lq
net\rq~ cooling time.)

Now consider the line indicated by ${\cal H}={\cal C}$, \ie the
equilibrium track, at high densities. Approaching this line from both
lower and higher temperatures we start from a region where $|t_{\rm
cool}| <t_{\rm H}$ and then pass onto the track where the denominator
of equation (\ref{eq:tc}) tends to zero very fast leading to an
infinite net cooling time, $t_{\rm
cool}\rightarrow\infty$. Consequently, we must go through a point where
$t_{\rm cool}=t_{\rm H}$ which explains why the solid line is basically
identical to the equilibrium track at sufficiently high densities. Gas
lying just below this line will be heated very quickly on a time scale
$\ll t_{\rm H}$ onto the track, and vice versa for gas slightly hotter
than the equilibrium $T$.

We can now understand the distribution of the gas in the ($\rho,T$)
plane as follows. Efficient photo-heating of under dense gas forces it
to remain at those temperatures where $t_{\rm cool}=t_{\rm H}$.  Where
gas falls into DM potential wells, shock heating <generates the large
plume of hot gas evident in the figure. Some of this gas may then reach
high enough densities so that cooling becomes efficient. This gas
condenses onto the equilibrium track where heating balances cooling.

\begin{figure}
\resizebox{\columnwidth}{!}{\includegraphics{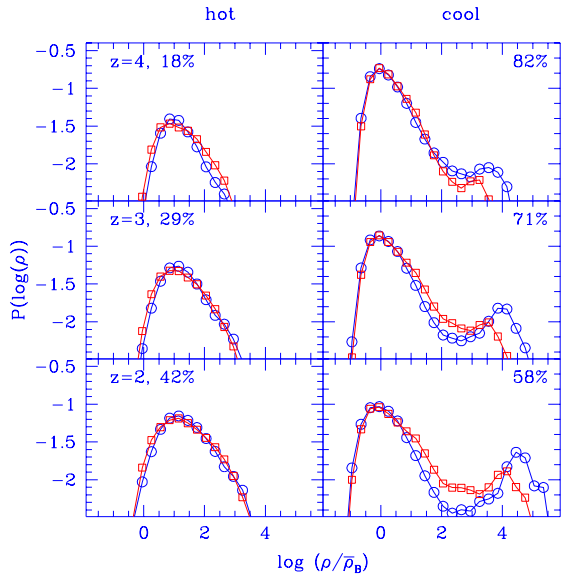}}
\caption{Mass fractions at given density in the hot and cool phases for
various redshifts for runs 22-64-k. Right hand panels refer to cool
gas, left hand panels to hot gas (see text), the redshifts are
indicated in the different panels.  \apm results are indicated by
circles, \hydra results by squares. Density is in units of the average
baryon density. The percentages in the panels refer to the respective
total mass fraction for \apm, which were very similar for \hydra.}
\label{fig:rhot}
\end{figure}

From the previous discussion it is clear that gas at both low and high
densities is forced by photo-heating to remain close to the line where
heating is dominant and provides a heating (net cooling) time equal to
the Hubble time (lower branch of the solid line in
figure~\ref{fig:figrt}). This line then defines a minimum temperature
at given density, $T_{\rm min}(\rho)$, for the simulated gas
distribution. Not evident from figure~\ref{fig:figrt} is that in fact
most of the particles lie very close to this minimum temperature. We
illustrate this in the following way. We label all gas with $T\le
1.5\times T_{\rm min}$ as being \lq cool\rq~ and the rest as being \lq
hot\rq~. The condition $T= 1.5\times T_{\rm min}$ is shown by the
dashed line in the figure. The distribution of cool and hot gas is
shown in detail in figure~\ref{fig:rhot} which compares the respective
mass fractions as a function of density for various redshifts. The
figure shows that at all redshifts most of the gas is in the cool
phase, though its fraction decreases with time. In addition we clearly
see an increasing amount of gas in the cool phase at the highest
densities, resulting from cooling in shocked halos at density
contrasts $\ge 200$.

Figure~\ref{fig:rhot} shows good agreement between \apm and \hydra for
the distribution of gas within each phase. The distribution of cool gas
at low densities (the largest mass fraction component) is very similar
between \apm \& \hydra, boding well for \lya simulations. This shows
that the low density correction in \hydra works well. The distribution
of hot gas is almost identical between the two codes as well, but there
are small differences in the high density cool phase: \apm halos have
less cool gas at intermediate densities, but more cool gas at the
highest densities, indicating that \apm halos tend to be more
concentrated at these highest densities. This is mostly due to small
differences in the gravitational softening employed in these runs (see
Table~\ref{table:runs}).

\subsubsection{Collapsed systems}
We have run a friends-of-friends group finder with linking length 0.177
times the average dark matter inter-particle distance on the output
files of A-22-64-k and H-22-64-k at $z=2.33$, thereby selecting halos
at an over density of $\sim 180$. Only halos with at least ten dark
matter particles are considered here. The centre of each dark matter
halo is defined by the position of the most strongly bound dark dark
matter particle. We then compute the spherically symmetric density
profile around the centre of the halo and remove all dark matter
particles further than $r_{200}$ from the centre, thus confining the
halo to its virialised part. Here, $r_{200}$ is the distance at which
the density falls below 200 times the mean density. In what follows,
SPH particles within $h+r_{200}$ from the centre of a halo are counted
as belonging to that halo.

\begin{figure}
\setlength{\unitlength}{1cm}
\centering
\begin{picture}(10,11)
\put(-1.,-4.0){\includegraphics{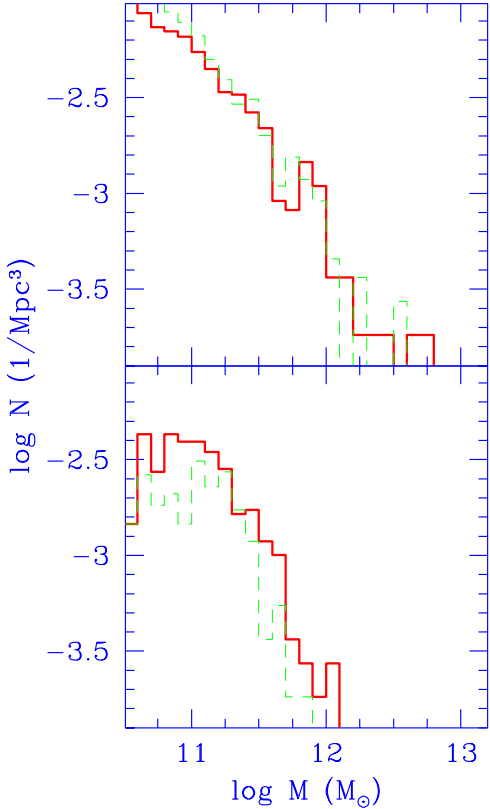}}
\end{picture}
\caption{Top panel: mass function for the simulation 22-64-k at
$z=2.33$ for \apm (full line) and \hydra (dashed line). $N$ denotes the
number of halos per comoving Mpc$^3$ with given mass in solar
units. At this resolution, a halo with mass $10^{11}~M_\odot$ contains
$\sim$ 37 dark matter particles. Bottom panel: same as top panel but
showing the mass function for the cooled gas in each halo. Cooled gas
masses are divided by the baryon fraction $\Omega_B=0.05$.}
\label{fig:massfunc}
\end{figure}

The mass function of the halos thus selected is illustrated in
figure~\ref{fig:massfunc} which shows the number $N$ of halos per
comoving volume as a function of virial mass. There is good agreement
between the two codes when comparing the total masses (\ie gas plus
dark matter) within the virial radius $r_{200}$. For each halo we have
also determined the amount of cooled gas, which for a halo with virial
temperature larger than $2T_4=2\times 10^4$K is gas with temperature
less than $2T_4$. As figure~\ref{fig:massfunc} shows, the \apm halos
tend to have slightly more cooled gas than the corresponding \hydra
ones. This is at least partly a consequence of the smaller
gravitational softening in the \apm simulations.

We have computed and compared several other quantities that
characterise the structure and dynamical properties of the halos, such
as their specific angular momentum and their shape parameters, as
determined from fitting the universal dark matter profiles of Navarro,
Frenk \& White (1996).  There is very good agreement between the two
codes on all these quantities.

\begin{figure}
\resizebox{\columnwidth}{!}{\includegraphics{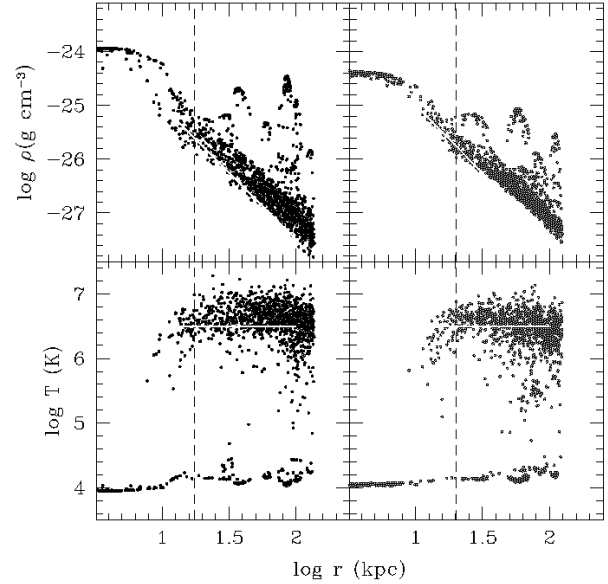}}
\caption{Comparison of gas density (top panels) and temperature (bottom
panels) between \apm (left panels) and \hydra (right panels) for a halo
of mass $5\times 10^{12}M_\odot$ in the 22-64-k simulations at
$z=2.33$. The \apm halo has higher density in the centre and also in
the in-falling satellites. Thick lines are the same in the left and
right panels and were drawn to guide the eye. The vertical dashed line
denotes the gravitational softening, the SPH smoothing length $h$ is at
least half that.}
\label{fig:halo1}
\end{figure}

\begin{figure}
\resizebox{\columnwidth}{!}{\includegraphics{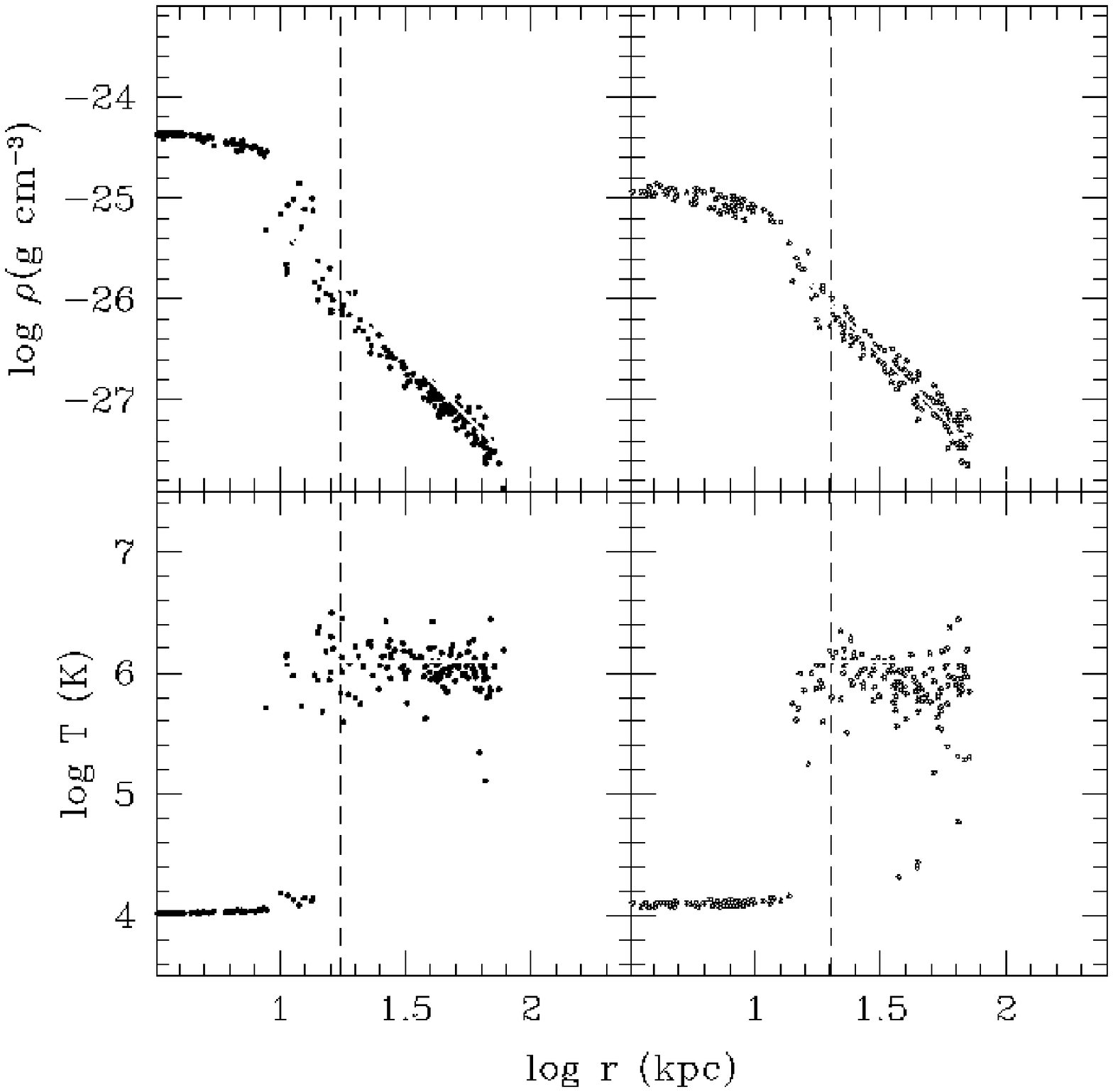}}
\caption{Same as figure~\ref{fig:halo1} but for a halo of mass
$5\times10^{11}M_\odot$. }
\label{fig:halo2}
\end{figure}

\begin{figure}
\resizebox{\columnwidth}{!}{\includegraphics{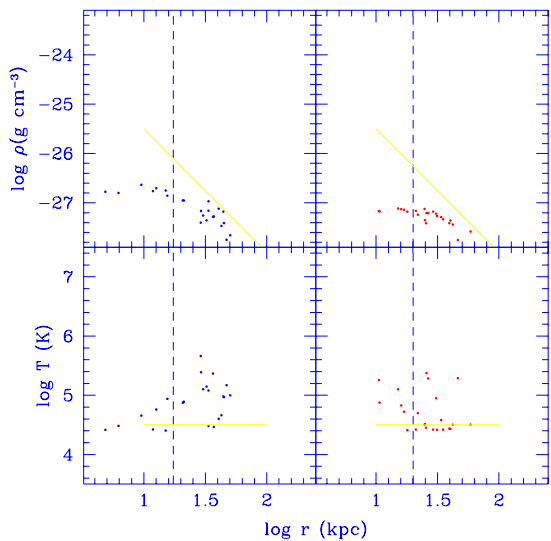}}
\caption{Same as figure~\ref{fig:halo1} but for a halo of mass
$5\times 10^{10}M_\odot$.}
\label{fig:halo3}
\end{figure}

We end this section by showing that the properties of the gas
distribution within halos are very similar as
well. Figures~(\ref{fig:halo1}-\ref{fig:halo3}) compare the density and
temperature distribution in three halos of total mass within the
virial radius of $5\times 10^{12}$, $5\times 10^{11}$ and $5\times
10^{10}M_\odot$ respectively. The mean and scatter in temperature as
well as density at any given radius is very similar in the two
codes. The \apm halos reach higher central densities because a smaller
gravitational softening is used. It is gratifying that the properties
of a halo with as few as $\sim 20$ particles of each type are
reasonably similar when comparing \apm with \hydra.


\subsection{Spectra}
To compare the statistics of simulated spectra with those presented by
Hernquist \etal (1996), we will use in this section both their initial
conditions and their cooling rates (our runs A-22-64-k and
H-22-64-k). These simulations should, in principle, use the spectral
evolution of Haardt \& Madau (1996) but with a reduced amplitude (see
Table~\ref{table:runs}), for which we give the fits to the ionising
spectrum in Appendix~\ref{sect:cooling}. Unfortunately, the actual
evolution of the background radiation is not identical, since there
were small changes between the fit to $J_{21}(z)$ in the preprint of
Haardt \& Madau (1996), from which Hernquist \etal took their
evolution, and the one which appeared in print (D. Weinberg, private
communication).

This Section is organised as follows. We begin by comparing spectra
along a given line of sight between \apm and \hydra. Next, we compare
global spectral measurements (mean absorption and one-point
distribution of the optical depth). Finally, we compare spectral
statistics (column- and $b$-parameter distributions) deduced from
automated Voigt-profile fitting of lines. Our detailed description of
the calculation of simulated spectra is given in
Section~\ref{sect:spectra}.

\subsubsection{Individual spectra}
\begin{figure}
\resizebox{\columnwidth}{!}{\includegraphics{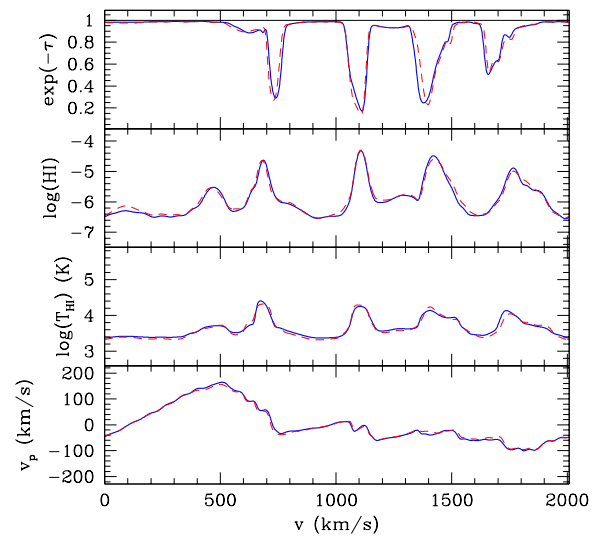}}
\caption{Absorption spectrum for \H through the middle of the box of
the 22-64-k simulation at $z=2.33$. From top to bottom: \H absorption
spectrum, \H fraction, \H weighted temperature, \H weighted peculiar
velocity. The top graph is versus velocity (wavelength), the bottom
three graphs are in units of position along the line of sight, from the
front of the box to the back. Solid lines refer to \apm simulation,
dashed lines to \hydra.}
\label{fig:Hspec}
\end{figure}

\begin{figure}
\resizebox{\columnwidth}{!}{\includegraphics{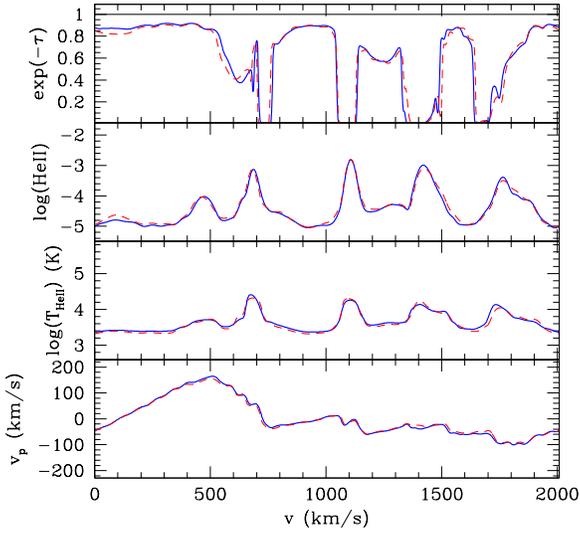}}
\caption{Same sight line as in figure~\ref{fig:Hspec} but now showing
\Hep results.}
\label{fig:Hepspec}
\end{figure}

\begin{figure}
\resizebox{\columnwidth}{!}{\includegraphics{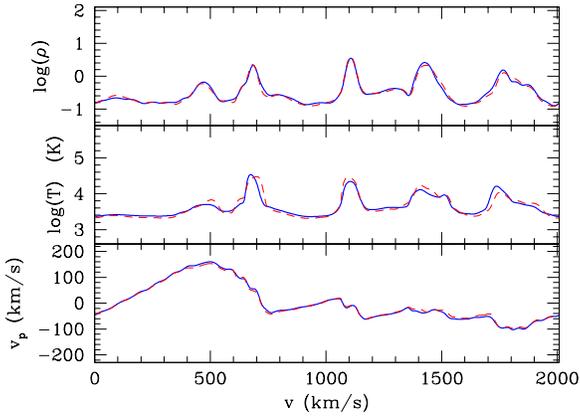}}
\caption{Comparison of the state of the IGM between \apm (full lines)
and \hydra (dashed line) along the same line of sight as
figures~\ref{fig:Hspec} and~\ref{fig:Hepspec}, showing gas density,
temperature and peculiar velocity (from top to bottom, respectively)
versus position in the box. Density is in units of the average gas
density at that redshift.}
\label{fig:Densspec}
\end{figure}

We have computed absorption spectra through the middle of the
computational box of the 22-64-k simulations at a redshift of
$z=2.33$. Figures~\ref{fig:Hspec}--\ref{fig:Densspec} compare the
results of \apm with \hydra, showing simulated spectra, density,
temperature and peculiar velocity along a particular line of
sight. These figures show that maxima in the density distribution give
rise to increased absorption with line widths determined by Doppler
broadening and Hubble expansion, as also shown by previous authors. For
hydrogen there are regions with very low absorption making a
distinction between \lq lines\rq~ and continuum possible, but this is
no longer true for helium nor for hydrogen at higher redshifts.  The
agreement between the two codes is impressive, both for \H and \Hep as
well as for the total density, over most of the velocity range. The
excellent agreement increases our confidence that both codes are
working properly and can be used to study the \lya forest.

\subsubsection{Global spectral characteristics}
\begin{figure}
\resizebox{\columnwidth}{!}{\includegraphics{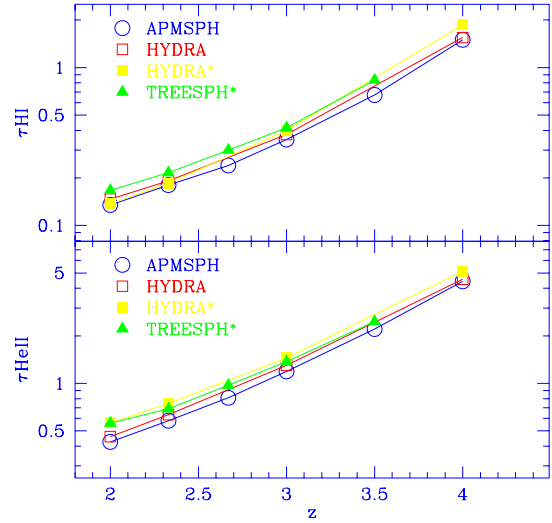}}
\caption{Mean effective optical depth $\tau$ as a function of redshift
$z$ for runs A-22-64-k \& H-22-64-k and published data from \tree.  The
top panel refers to hydrogen, the bottom panel to singly ionised
helium. The \apm \& \hydra curves are computed from $32^2$ sight lines
through the box. The \tree data were taken from figure~ 3 in Croft
\etal (1997). The \hydrastar and \treestar use the fit equations.~3-5
from Croft \etal (1997) for the evolution of the ionization rate
divided by two, the other two use the fit from
table~\ref{table:Haardt}, divided by two.}
\label{fig:meanabsorption}
\end{figure}

\begin{figure}
\resizebox{\columnwidth}{!}{\includegraphics{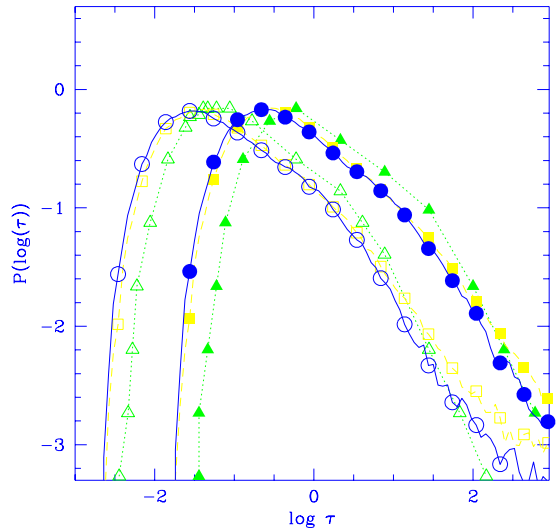}}
\caption{One-point probability distribution of optical depth for \H
(open symbols) and \Hep (filled symbols) for the same simulation as
shown in figure~\ref{fig:meanabsorption} at a redshift of
$z=2.33$. Symbols: \apm: circles; \hydra: squares; \tree:
triangles. The \apm and \hydra curves assume the fit in
Table~\ref{table:Haardt} for the background radiation; the \tree data
were taken from figure~11 in Croft \etal 1997.}
\label{fig:ptau}
\end{figure}

\begin{figure}
\resizebox{\columnwidth}{!}{\includegraphics{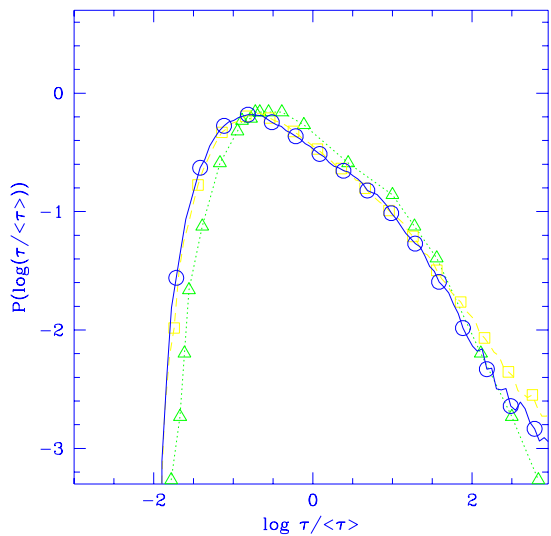}}
\caption{Same as figure~\ref{fig:ptau} but for probability distribution
of the \H optical depth in units of the mean.}
\label{fig:ptau_av}
\end{figure}

We have computed simulated spectra along 1024 lines-of-sight on a
square grid through the box of the 22-64-k simulations at various
redshifts. From the spectra we have computed the effective mean optical
depth at this redshift, $\bar\tau(z)\equiv -{\rm ln} (\sum_1^{N_p}
e^{-\tau}/N_p)$ where the sum is over $N_p$ pixels of the
spectrum. Figure~\ref{fig:meanabsorption} shows that \apm, \hydra and
\tree predict a very similar redshift evolution for $\bar\tau(z)$. Most
of the remaining differences are presumably due to the slightly
different assumed evolution of the background flux, as can be seen by
comparing the \hydrastar and \treestar curves. The latter runs use
identical fits to the photo-ionising flux although they still use a
different fit to the photo-heating rates. A more detailed comparison is
shown in figure~\ref{fig:ptau} which depicts the one-point probability
distribution for the optical depth in a pixel. Again the comparison is
satisfactory, with most of the difference between the Tree and the P3M
codes being due to the different background flux. Note that the
apparent shift in the mean of $P(\tau)$ between Tree and P3M codes is
not just due to the difference in $\bar\tau$. Indeed, when the
distribution of $P(\tau/\bar \tau)$ is plotted as in
figure~\ref{fig:ptau_av} then some differences remain, indicating that
also the shape of $P(\tau)$ is not quite the same between the
codes. The \apm \& \hydra codes show small differences in the high
$\tau$ tail, which is a consequence of the small differences in the
properties of the high density gas, as we remarked on earlier.

\subsubsection{Voigt-profile analysis}
In the previous sections we have shown that \apm and \hydra spectra
along a given line of sight look very similar and that global
statistical quantities of the optical depth distribution are similar as
well. Here we want to compare the statistical properties of the
simulated absorption lines based on Voigt profile (VP) fitting. VP
fitting has consistently been used to analyse the complex blended
features of the \lya forest, seen in real quasar spectra (\eg Webb
1987, Carswell \etal 1987) and in numerical simulations (Dav\'e \etal
1997, Haehnelt \& Steinmetz 1998, Zhang \etal 1998). The technique was
devised initially under the premise that the forest lines arose from
discrete absorbing clouds intervening the quasar line of sight, a
picture currently challenged by the successful reproduction of such
observed VP parameter distributions by the smoothly distributed IGM
naturally occurring within gas-dynamical simulations.

As we have seen earlier, many of the \lq lines\rq~ occurring in these
simulations have significant distortions from the simple Voigt profile
due to the fact that the structure giving rise to the line is extended
in space and thus still takes part in the Hubble expansion. VP-fitting
of such a system then requires a complex blend of several Voigt
profiles with a range in column density and velocity
width. Consequently, the parameters characterising these components
cannot automatically be used to infer physical properties of the
absorbers. Given the loss of a physical justification for the VP
approach, the method serves merely as way of characterising the
undulations of what has been described more accurately by authors as a
\lya Gunn-Peterson-effect (Gunn \& Peterson 1965). Its power as a
diagnostic tool for doing this must be scrutinised, before agreement
with the observations can be described as a success for the model.
However, since we want to compare our simulated spectra with observed
ones, it is important that we analyse them in the same way. We
therefore use a standard package for VP fitting (Carswell \etal 1987)
used frequently by observers. We will also show that the results
deduced from VP-fitting are relatively insensitive to the detailed
implementation of the profile fitting. There is however one remaining
difference between the simulated and real spectra, namely our simulated
spectra are not superposed onto the continuum of a background
quasar. It is therefore difficult to model the continuum fitting
performed by observers. This problem is aggravated by the small length
of our simulated spectra compared to observed ones. We try to estimate
the uncertainty introduced by this below.

In this paper, we use an automated version of the VPFIT software
(Carswell \etal 1987) written to perform $\chi^2$ minimisation over
many VP parameters given observational spectra. Prior to using VPFIT,
each spectrum is convolved with a Gaussian profile with FWHM = 8 km
s$^{-1}$, then re-sampled onto pixels of width 3 km s$^{-1}$ to mimic
the instrumental profile and characteristics of the HIRES spectrograph
on the Keck telescope, which currently provides the most up-to-date
results on the properties of the weak \lya forest lines. Artificial
noise is introduced by adding a Gaussian random signal with zero mean,
and standard deviation $\sigma=0.02$ to every pixel (\ie a SNR of 50
for pixels at the continuum), mimicking the read-out noise dominated
character of modern observed spectra.

In our simulated spectra, zero-order absorption occurs across each line
of sight that would presumably be removed during observational analysis
by the continuum fitting procedure. In analysing real spectra, the
continuum is normally determined by fitting a low-order polynomial to
apparently \lq unabsorbed\rq~ regions of the spectrum that are
typically much longer than our simulated spectra. This difference in
spectral range means that we cannot mimic the continuum fitting
procedure used by observers on our simulated spectra. The observational
procedure depends a great deal on the quality of data as well as on
redshift since finding \lq unabsorbed\rq~ regions of the spectrum
becomes more difficult at higher redshift, where the \lya forest
opacities are higher. Our spectra cover a small enough velocity range
to be fit by a flat continuum, as chosen by a simple procedure
described as follows. A low average continuum level is assumed
initially, then all pixels below and not within 1 $\sigma$ of this
level are rejected, and a new average flux level for the remaining
pixels is computed. The same condition is applied for this new level
and so-on, until the average flux varies by less than 1\% (note that
the signal to noise ratio adopted is 50 at the continuum). This final
average flux level is adopted as the fitted continuum, and the spectrum
is renormalised accordingly.  We test for the effects of varying this
level on the VP results (see below).

The final stage in preparing spectra for VPFIT is via an automatic
procedure to find first guess profiles for each line. We make initial
guesses using the method of Dav\'e \etal (1997) for finding weak lines,
\ie firstly smoothing the spectrum before \lq growing\rq~ a Voigt
profile into the deepest depression. We repeat this procedure on the
spectrum with the current best fit Voigt profile subtracted, until the
residual absorption varies by less than 20\%.  VPFIT turns out to be
robust in finding accurate fits given few and even quite wrong initial
guesses, so we are confident that our results do not depend strongly on
the details of the procedure for obtaining first guesses. VPFIT then
uses these guesses to find the best-fitting profiles, adding more lines
and removing ill-constrained lines where necessary. In general, we
calculated VP parameters for a spectrum until an overall reduced
$\chi^2$ value for the fit falls below 1.3, unless otherwise
indicated. This is carried out for typically 300 sight line spectra
taken on a grid through each simulation. We check below the effects of
varying the required minimum reduced $\chi^2$ value.

It is not clear {\it a priori} that the deduced VP statistics are
independent of the fitting procedure, given \eg the non-uniqueness of
Voigt profiles in fitting a blended line and the fact that observers
often fit profiles by hand. Nevertheless, we will show below that our
deduced statistics are in excellent agreement with the ones obtained by
Dav\'e \etal (1997), using an independent fitting procedure.

\begin{figure}
\resizebox{\columnwidth}{!}{\includegraphics{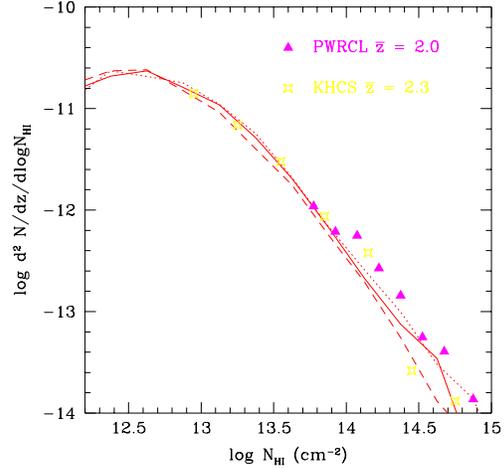}}
\caption{Column density distributions, $d^2N/dz/dlog N_\h$, from an
analysis using VPFIT of 300 spectra along lines of sight through
several 22Mpc, $64^3$ particle simulations, at $z=2$. Solid and dashed
lines show results of A-22-64-k, and H-22-64-k simulations
respectively, run using the same initial conditions as used by
Hernquist \etal (1996) (c.f. figure~2 of Dav\'e \etal 1997). The dotted
line shows results of an \apm simulation run with different initial
conditions (A-22-64). Observational results of PWRCL and KHCS with mean
absorption redshifts $\bar{z} = 2.0$ and 2.3 respectively, are also
plotted as indicated.}
\label{fig:columndens2}
\end{figure}

\begin{figure}
\resizebox{\columnwidth}{!}{\includegraphics{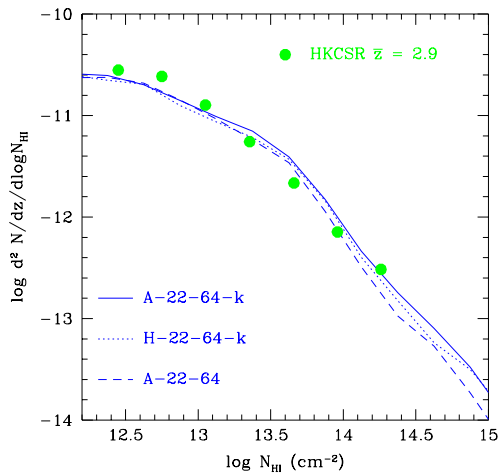}}
\caption{Same as figure~\ref{fig:columndens2} but at
$z=3$. Observational results of HKCSR with mean absorption redshifts
$\bar{z} = 2.9$ are plotted.}
\label{fig:columndens3}
\end{figure}

In figure~\ref{fig:columndens2} we show the column density distribution
(the number $N$ of lines per unit redshift and column density
$d^2N/dz/d\log N_\h$, or \lq differential distribution function\rq~,
DDF) for runs 22-64-k which use the initial conditions from Hernquist
\etal (1996) undertaken using \apm (solid), and \hydra (dotted) at
redshift 2. Also shown are the results of a second similar simulation
using \apm (dashed), but where the initial conditions are drawn from a
different random realization of the initial density field (run
A-22-64). All three simulations produce very similar DDFs across the
range of column densities $\le 10^{15}$ cm$^{-2}$. There is good
agreement between the two codes at $z=3$ as well, as illustrated in
figure~\ref{fig:columndens3}. It is gratifying to conclude that these
results should not depend greatly on either our code implementations,
nor the specific realization chosen, nor the VP-fitting procedure as
applied to two different sets of spectra at both redshifts.

Superposed on the simulated DDFs are observationally determined points
of Petitjean \etal (1993, hereafter PWRCL) and Kim \etal (1995,
hereafter KHCS) for redshift $z=2.0$ and 2.3, and Hu \etal (1995,
hereafter HKCSR), for $z=3$. We see that the simulations follow the
observed points well, though possibly under predicting lines with $N_\h
> 10^{13}$ cm$^{-2}$ at $z=2$. More importantly however, our simulated
DDFs are directly comparable to the derived column density
distributions of figures~2 and 3 of Dav\'e \etal (1997) where the same
cube simulated using \tree was analysed in a similar way using the
AUTOVP/PROFIT fitting software, instead of VPFIT. We find that there is
a very close similarity. The results of Dav\'e \etal at $z=2$ also
under predict the points of PWRCL and have a very similar slope to our
DDF for $N_{\h}> 10^{13}$ cm$^{-2}$. Further at $z=3$, the distribution
lies constantly above the $z=2$ results for $N_{\h} > 10^{13.5}$
cm$^{-2}$, but has a shallower slope at lower column densities. The
exact comparison for $N_{\h} < 10^{13.5}$ cm$^{-2}$, as shown later, is
sensitive to the VP-fitting conditions, nevertheless the similarity is
striking enough to suggest that the detailed nature of the weak \lya
forest column density distributions are reproducible given both
different simulation codes and VP-fitting software.

\begin{figure}
\resizebox{\columnwidth}{!}{\includegraphics{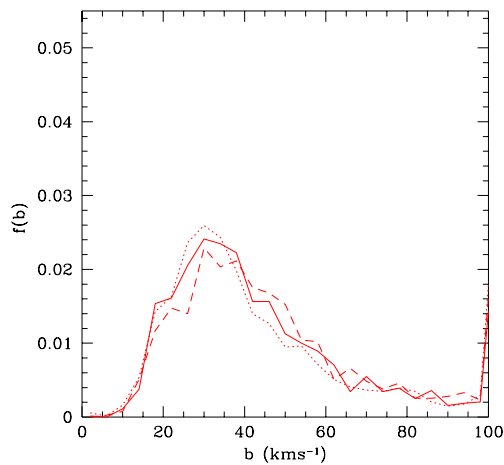}}
\caption{$b$-parameter distributions correspondoing to the column
densities shown in figure~\ref{fig:columndens2}. Only lines with $N(\H)
> 10^{13}$ cm$^{-2}$ are included.}
\label{fig:bpar2}
\end{figure}

\begin{figure}
\resizebox{\columnwidth}{!}{\includegraphics{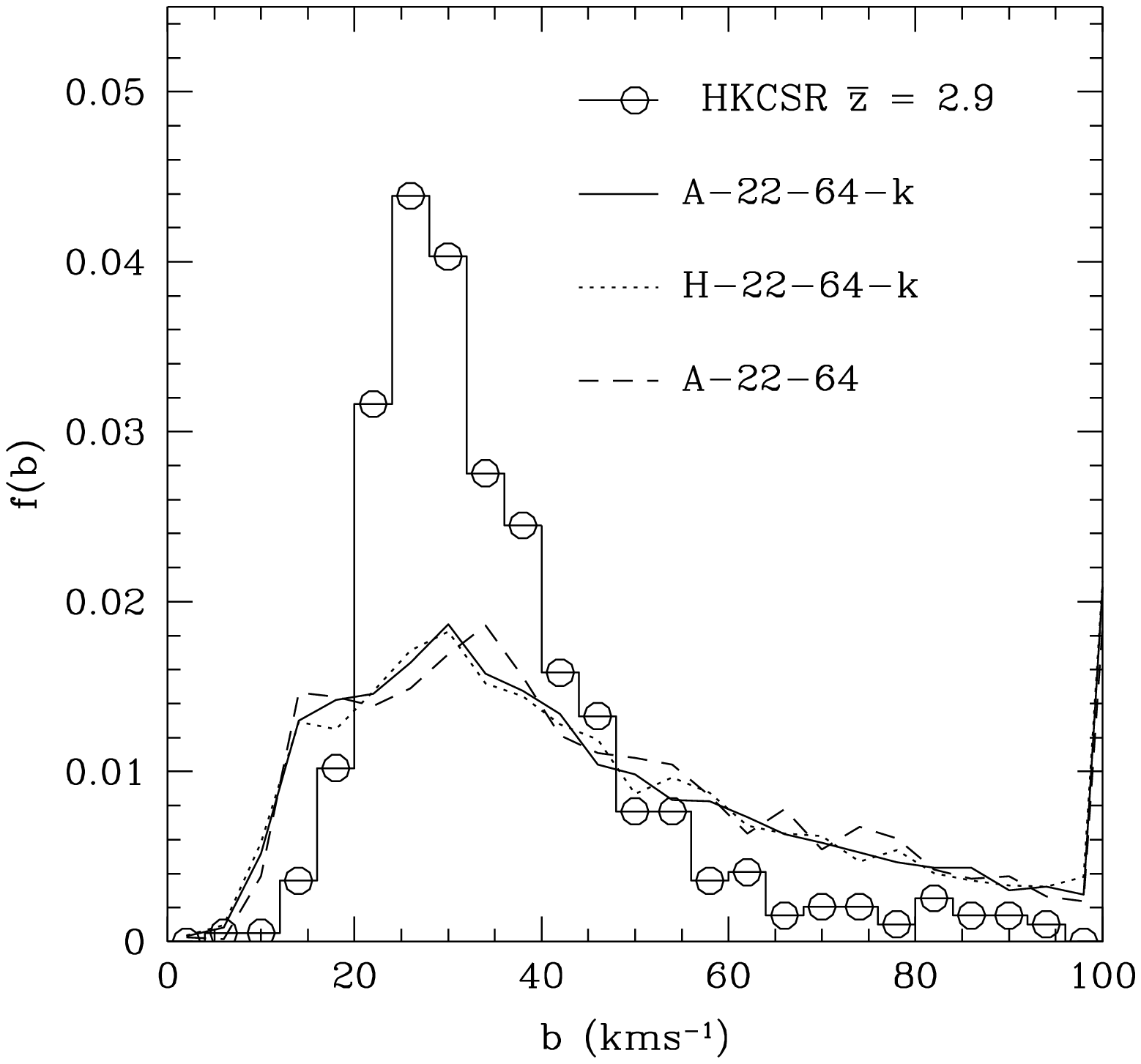}}
\caption{$b$-parameter distributions corresponding to the column
densities shown in figure~\ref{fig:columndens3}. Only lines with $N(\H)
> 10^{13}$ cm$^{-2}$ are included. The histogram shows the corresponding
observational results of HKCSR.}
\label{fig:bpar3}
\end{figure}

Figures~(\ref{fig:bpar2}-\ref{fig:bpar3}) show the derived
$b$-parameter distributions for each simulation described above. We
follow Dav\'e \etal (1997) and include only lines with $N_\h > 10^{13}$
cm$^{-2}$ to eliminate broad, weak features, which are sensitive to the
VP-fitting process, from biasing the comparison with
observations. These distributions show a consistency indicating
independence with simulation code and realization. The shape of the $b$
parameter distribution is very similar to the one obtained by Dav\'e
\etal (1997, see their figure~3), from their analysis of \tree
simulations with the same initial conditions and resolutions as our
A-22-64-k and H-22-64-k runs.  However there is a clear discrepancy
between the observed $b$ distribution as determined by HKCSR at $z=3$
from observations and the simulation results at that redshift. Our
simulations all over-predict lines with $b<20$ and $b>50$ km s$^{-1}$,
whilst failing to reproduce the observed high peak of lines at $b \sim
30$ km s$^{-1}$. Although the $b$ parameter distribution is sensitive
to the signal to noise ratio and the VP set-up, we will show below that
the dominant reason for the discrepancy with observations is lack of
numerical resolution.

\section{Numerical effects}
\label{sect:resolution}
In the previous section we have shown that there is excellent agreement
between \apm, \hydra and \tree on a variety of statistics that
characterise the simulated \lya spectra. The different codes, however,
were compared at identical numerical resolution. In this section, we
will investigate how sensitive these statistics are to details of the
simulation and analysis, such as numerical resolution, box size, and
details of the VP-fitting process. To investigate the effects of
numerical resolution one would ideally like to do a simulation of a
given region of space at several resolutions to asses the degree of
convergence. However, since the required CPU time to perform a given
simulation rapidly increases with the number of particles in the run,
we have opted to keep the number of particles fixed but decrease the
box size. This complicates the analysis since differences in statistics
between a small and a large box may be due to the lack of long
wavelength perturbations and saturation of modes in the smaller box,
rather than non-convergence of the simulation in the larger box. In the
same vein as the previous section we will describe first how global
properties depend on resolution before concentrating on line
statistics. We will then investigate the dependence of line statistics
on the VP-fitting procedure. Finally, we will investigate how line
statistics depend on the assumed UV background.

\subsection{Numerical resolution}
\subsubsection{Global spectral characteristics}
\begin{figure}
\resizebox{\columnwidth}{!}{\includegraphics{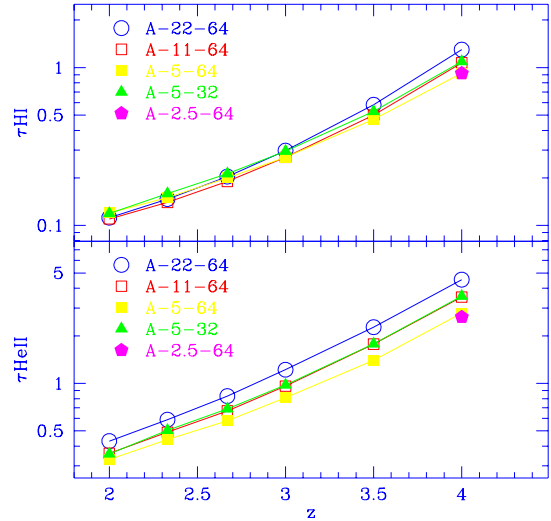}}
\caption{Mean effective optical depth $\bar\tau$ as a function of
redshift $z$ for several box sizes indicated in the figure. The top
panel refers to hydrogen, the bottom panel to singly ionised
helium. The $64^3$ simulations have mass resolutions of $1.4\times
10^8$, $1.8\times 10^7$, $2.2\times 10^6$ and $2.1\times 10^5M_\odot$
per SPH particle, respectively.}
\label{fig:meanabsorptionres}
\end{figure}

We have computed the effective mean optical depth $\bar\tau(z)$ at
various redshifts as well as the one-point probability distribution
$P(\tau)$ from 1024 lines-of-sight for the A-22-64, A-11-64 and A-5-64
simulations. Figure~\ref{fig:meanabsorptionres} shows only small
differences for the $\bar\tau_\h$ optical depth as the resolution is
increased. Note that the optical depth tends to {\em decrease} with
increasing resolution. The main reason for this is that at higher
resolution, small halos collapse that were not resolved at lower
resolution. This decreases the density of gas in the low density
regions which leads to a decreasing optical depth. This effect is much
more pronounced for helium, for which there are still significant
differences in mean optical depth between A-11-64 and A-5-64. Note that
this is a resolution and not a box size effect: the A-5-32 box also
shown in the figure has the same resolution as the A-11-64 box and
gives the same optical depths despite its different box size. We have
therefore run an even higher resolution simulation, A-2.5-64, which
gives a mean optical depth for \Hep at $z=4$ of $\bar\tau=2.63$ which
is reasonably close to the value 2.78 of the A-5-64 simulation. This
suggests that the \Hep absorption has almost converged in our highest
resolution simulations, but not in the lower resolution ones.

\begin{figure}
\resizebox{\columnwidth}{!}{\includegraphics{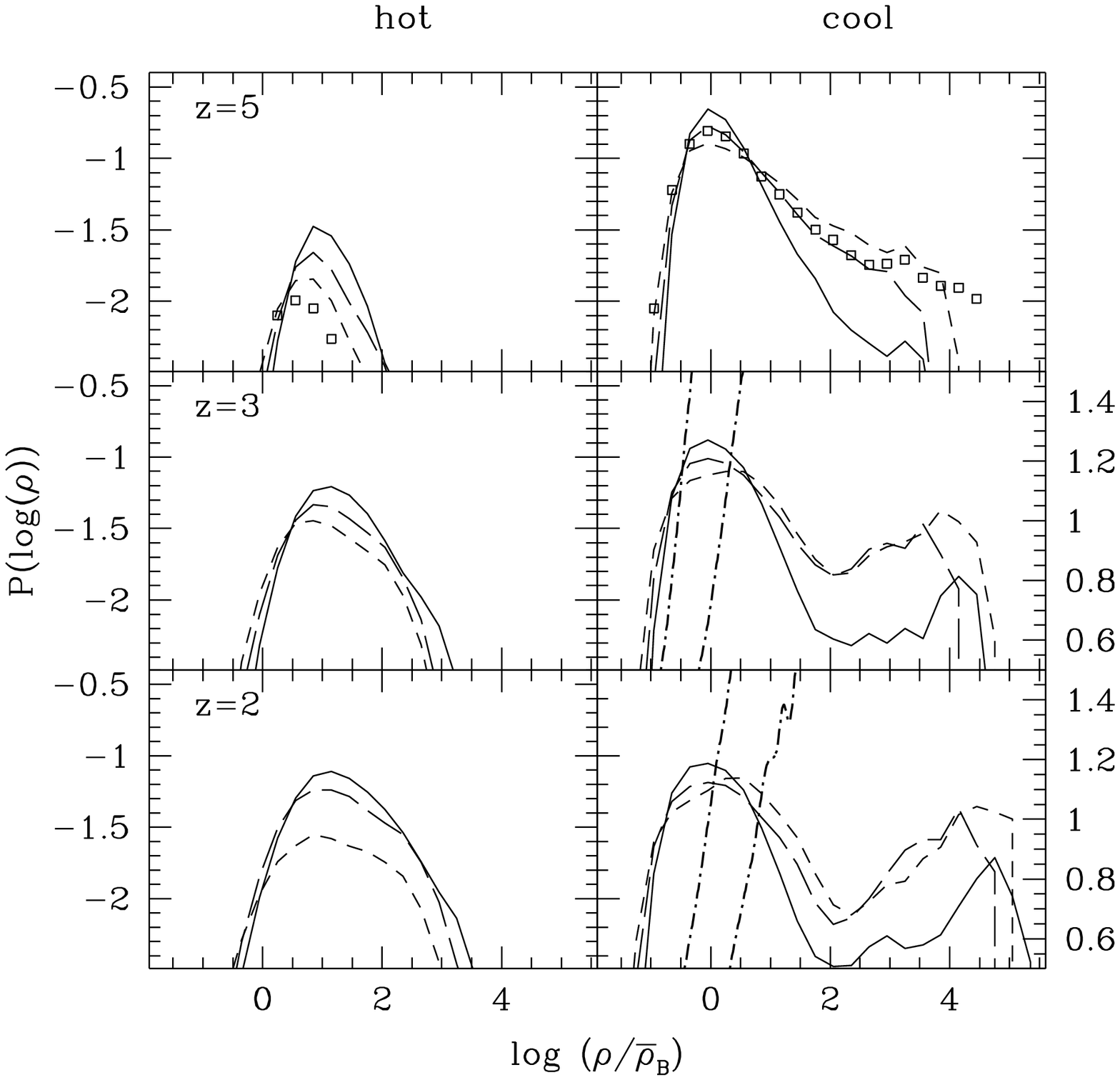}}
\caption{}{Mass fractions at given density in the hot and cool phases
for various redshifts for runs A-22-64 (full lines), A-11-64 (long
dashed) and A-5-64 (short dashed) for various redshifts as indicated in
the panels (see Section~\ref{sect:globalgas} for definition of
cool/hot). At higher resolution a larger fraction of gas collapses to
higher densities leading to less gas at densities $\rho\sim
(0.1-0.3)\times \bar{\rho}_B$ which contributes most to the
absorption. The symbols in the $z=5$ panels refer to our highest
resolution A-2.5-64 run and compare well with the lower resolution
A-5-64 run. The dot-dashed lines in the cool gas panels at $z=3$ and
$z=2$ indicate the effective mean optical depth $\bar{\tau}(\rho)$ as a
function of density measured from the simulations for hydrogen and
helium (curve giving higher values, right scale).}
\label{fig:rtres}
\end{figure}

Zhang \etal (1997) already pointed out that it is far more difficult to
obtain convergence for the effective mean optical depth for helium than
for hydrogen. To explain the reason for this we first define the
effective mean optical depth $\bar{\tau}(\rho)$ at given density $\rho$
from
\begin{equation}
\exp(-\bar{\tau}(\rho)) = \sum \exp\left(-\tau(\rho)\right)/N(\rho)\,,
\label{eq:meantau}
\end{equation}
where the sum is over all $N(\rho)$ pixels in the simulated spectra
where the {\em real} space density in the particular pixel is $\rho$
(see \eg the discussion of figures~6c \& 6d in Croft \etal 1997). For
later use we define the normalised volume fraction of gas at density
$\rho$ as $P(\rho)=N(\rho)/N_p$, where $N_p$ is the total number of
pixels. Now, at higher resolution a larger fraction of gas collapses
into halos with a modest over density which are simply not resolved in
lower resolution simulations. At $z=2$ for example, this reduces the
fraction of gas at densities $\sim (0.1-1)\times \bar{\rho}_B$ and
correspondingly increases that fraction at densities $\sim (1-10)\times
\bar{\rho}_B$, as is illustrated in figure~\ref{fig:rtres}, where we
show the distribution of cool and hot gas (as defined in
Section~\ref{sect:globalgas}) at each resolution. The gas that
collapses to higher densities has an effective mean optical depth
$\simlt 1$ in hydrogen yet $\tau\gg 1$ for helium, as is also shown in
the figure. Therefore, for hydrogen some of the absorption lost because
of the decreased fraction of low density gas is recuperated from the
increased opacity in the higher density gas. This is not true for
helium, however, since the lower density gas is already optically thick
and increasing its column density does not significantly change its net
absorption. The same reasoning applies at higher redshifts where it is
even clearer that the amount of cool gas is resolution dependent and
has only just converged in our highest resolution
simulation. Miralda-Escud\'e (1993) previously remarked that for a
large jump in ionising flux from the \H limit to the \Hep limit, as
occurs in the Haardt \& Madau (1996) spectra, lines with an \H column
density of $\sim 10^{12}$ cm$^{-2}$ have central \Hep optical depths
$\sim 1$. Hence for a simulation to have a reliable \Hep mean effective
optical depth it needs to resolve well lines with \H column density of
$\ltsima 10^{12}$ cm$^{-2}$. We will show later
(figure~\ref{fig:Nnumer2}) that the A-22-64 simulation produces far
fewer lines with such low column densities than the higher resolution
runs, which then explains why the low resolution simulation has a
considerably higher mean effective \Hep optical depth.

We have shown in figure~\ref{fig:rtres} that the amount of cool gas is
strongly resolution dependent. In particular, run 22-64-k, which has
identical parameters and resolution to the Hernquist \etal (1996)
simulation, underpredicts the amount of cool, high density gas by a
large fraction (a factor of $\sim 5$ at $z=3$). Since lines of sight
intersecting high density cool gas regions produce damped \lya
absorbers, the statistics of those damped \lya systems, as deduced from
these simulations (Katz \etal 1996a), are uncertain because they are
very sensitive to numerical resolution.

\begin{figure}
\setlength{\unitlength}{1cm}
\centering
\begin{picture}(10,11)
\put(-1., -4.0){\includegraphics{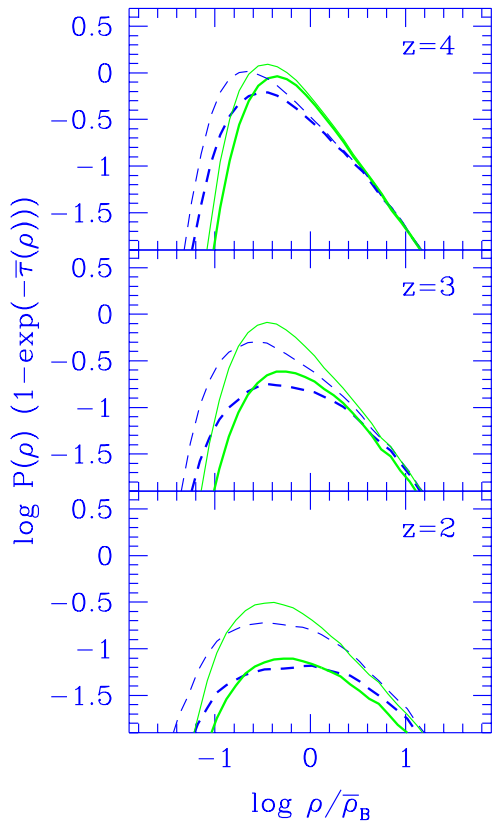}}
\end{picture}
\caption{Mean transmission at given density, $1-\exp(-\bar{\tau}(\rho))$,
weighted by the fraction by volume of pixels $P(\rho)$ at that density,
plotted versus density, illustrating which densities contribute most to
the absorption as a function of redshift, indicated in the panels. Full
curves: A-22-64, dashed curves: A-5-64, for hydrogen (thick lines) and
helium (thin lines).}
\label{fig:ptau_av_res}
\end{figure}

The influence of resolution on optical depth is further illustrated in
figure~\ref{fig:ptau_av_res} which compares the net transmission of gas
as a function of its density for our highest and lowest resolution
runs. The figure illustrates clearly that at higher resolution the
transmission in the low density gas increases significantly but for
hydrogen this is mostly compensated by a decrease in transmission in
the higher density pixels. This compensation does not happen for helium
and consequently the helium optical depth decreases noticeably with
increased resolution. Note also that, although the hydrogen mean
optical depth has converged there are still significant differences in
the distribution of $P(\rho)\, (1-\exp(-\bar{\tau}(\rho)))$.

\subsubsection{Line statistics}
\begin{figure}
\resizebox{\columnwidth}{!}{\includegraphics{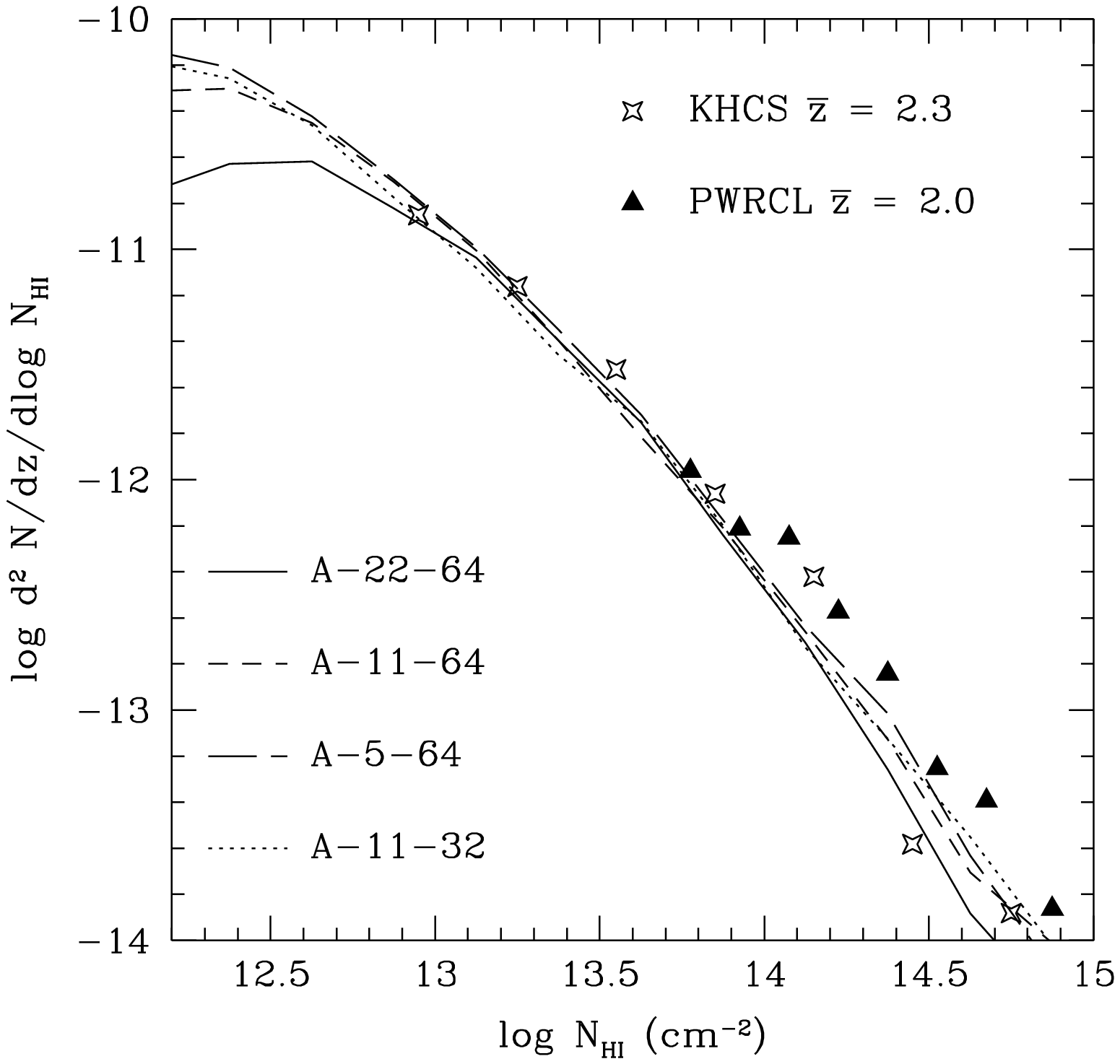}}
\caption{Column density distribution dependence on box-size and
resolution at z=2. Results are shown for three simulations run using
\apm with the same initial modes, but with box size and particle
numbers as indicated. Observational results of KHCS and PWRCL are also
plotted as indicated.}
\label{fig:Nnumer2}
\end{figure}

\begin{figure}
\resizebox{\columnwidth}{!}{\includegraphics{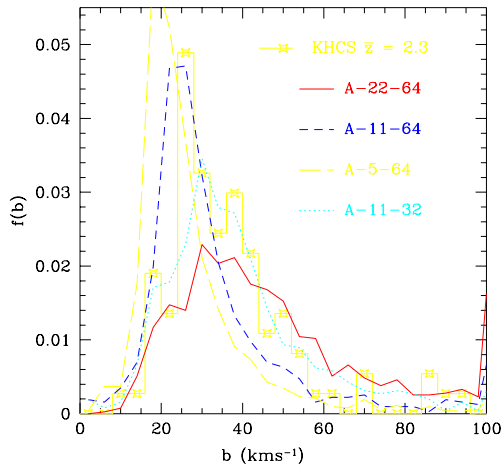}}
\caption{Corresponding $b$ parameter distributions to the column
densities shown in figure~\ref{fig:Nnumer2} for lines with $N_h > 13$
cm$^{-2}$, at $z=2$. The histogram shows the corresponding
observational results of KHCS. Note that the highest bin plotted for
the KHCS results contains only 18 lines.}
\label{fig:bnumer2}
\end{figure}

\begin{figure}
\resizebox{\columnwidth}{!}{\includegraphics{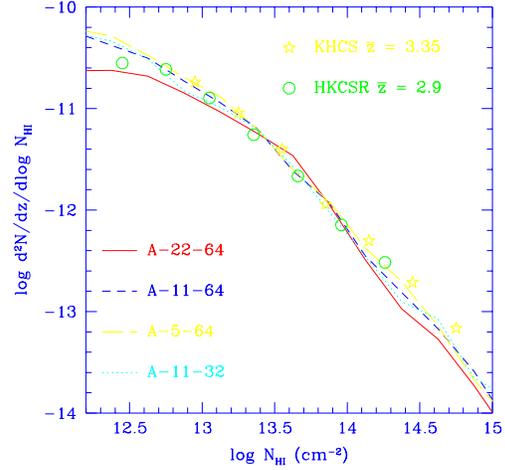}}
\caption{Same as figure~\ref{fig:Nnumer2} for $z=3$. Observational
results are taken from KHCS and HKCSR, using independent data.}
\label{fig:Nnumer3}
\end{figure}

\begin{figure}
\resizebox{\columnwidth}{!}{\includegraphics{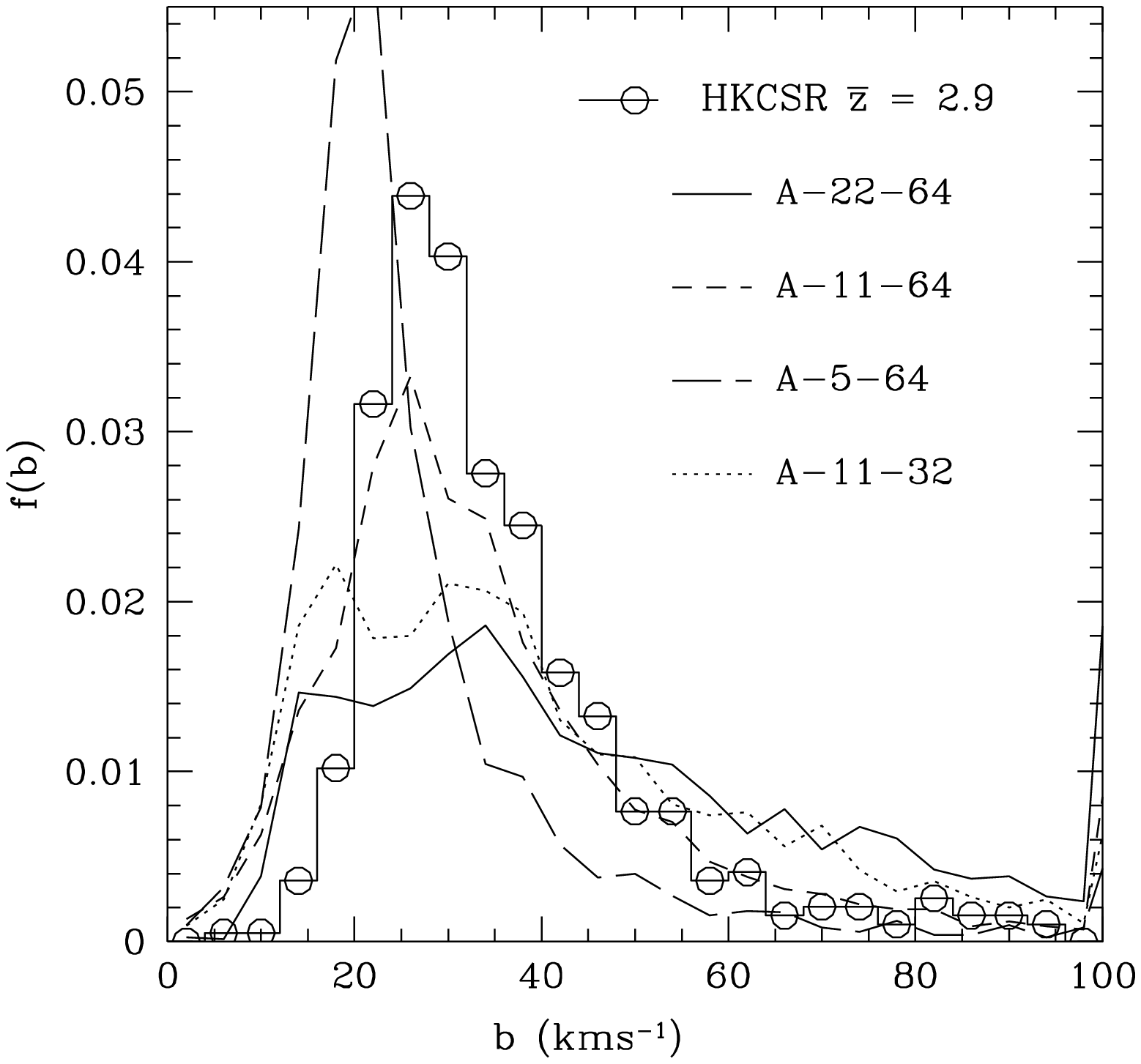}}
\caption{Same as figure~\ref{fig:bnumer2} for $z=3$. The histogram
shows the observational results of HKCSR.}
\label{fig:bnumer3}
\end{figure}
We have used VP-fitting with the set-up as described earlier to
characterise the absorption lines in simulated spectra to investigate
the extent to which VP distributions depend on numerical
resolution. The results are shown in
figures~\ref{fig:Nnumer2}-\ref{fig:bnumer3} for redshifts 2 and 3. The
number of hydrogen lines with column density $\ge 10^{13}$ cm$^{-2}$ is
largely independent of resolution and box size at both redshifts, as
shown by the DDFs in figures~\ref{fig:Nnumer2} and \ref{fig:Nnumer3},
which is quite encouraging. However below this column density, the
lower resolution simulation A-22-64 starts to lose a significant number
of lines. The higher resolution simulations show excellent agreement
with the observed DDFs also indicated in the figures for both
redshifts, although there may be a hint that the simulations
underproduce lines at redshift 2.

This agreement does not extent to the $b$-parameter distributions
however. Simulations at different resolutions produce quite different
$b$-parameter distributions, none of which fit particularly well the
observed one. Higher resolution simulations produce a larger fraction
of narrower lines. Note that the $b$-parameter distributions shown in
the figures are restricted to lines with $N_\h\ge 10^{13}$ cm$^{-2}$
where there was good agreement for the DDF between simulations at
different resolution. We will show in the next section that these
$b$-parameter distributions are in fact also quite sensitive to the
parameters used in the VP-fit analyses (\eg required reduced $\chi^2$
and continuum placement), complicating the comparison with
observational data.

\subsection{Dependence on VP-fitting procedure}
\begin{figure}
\resizebox{\columnwidth}{!}{\includegraphics{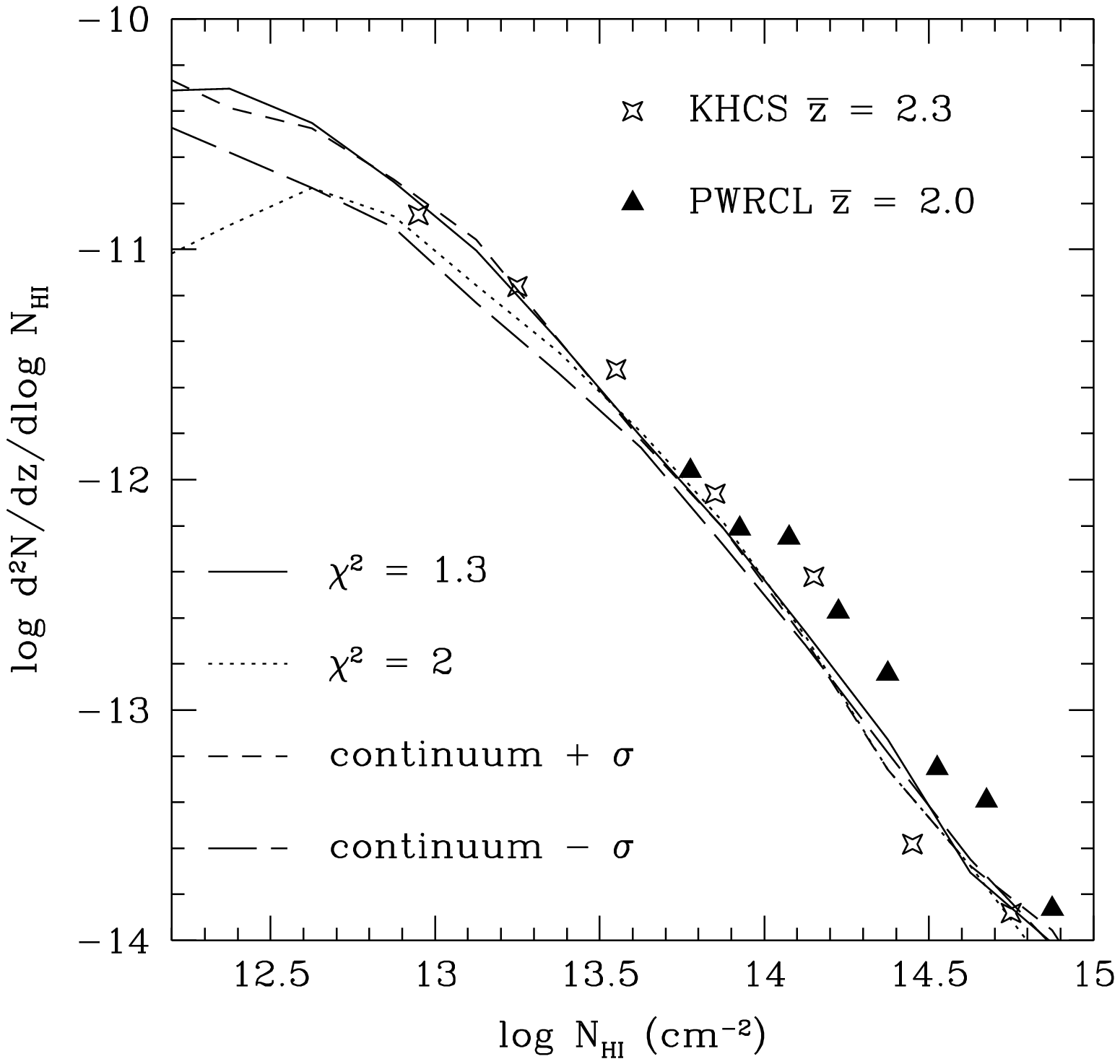}}
\caption{Column density distributions derived from the same A-11-64
simulation at $z=2$ but varying the reduced $\chi^2$ requirement and
continuum level chosen for each spectrum. Solid and dotted lines show
results requiring the overall VP-fitting to each spectrum to have a
reduced $\chi^2 < 1.3$ and 2 respectively. Short-dashed and long-dashed
lines show the effect of respectively raising and lowering the chosen
continuum fit for each spectrum by 2\%, prior to VP-fitting, for
$\chi^2 < 1.3$. Observational results of KHCS and PWRCL are also
plotted as indicated.}
\label{fig:Nobs}
\end{figure}

\begin{figure}
\resizebox{\columnwidth}{!}{\includegraphics{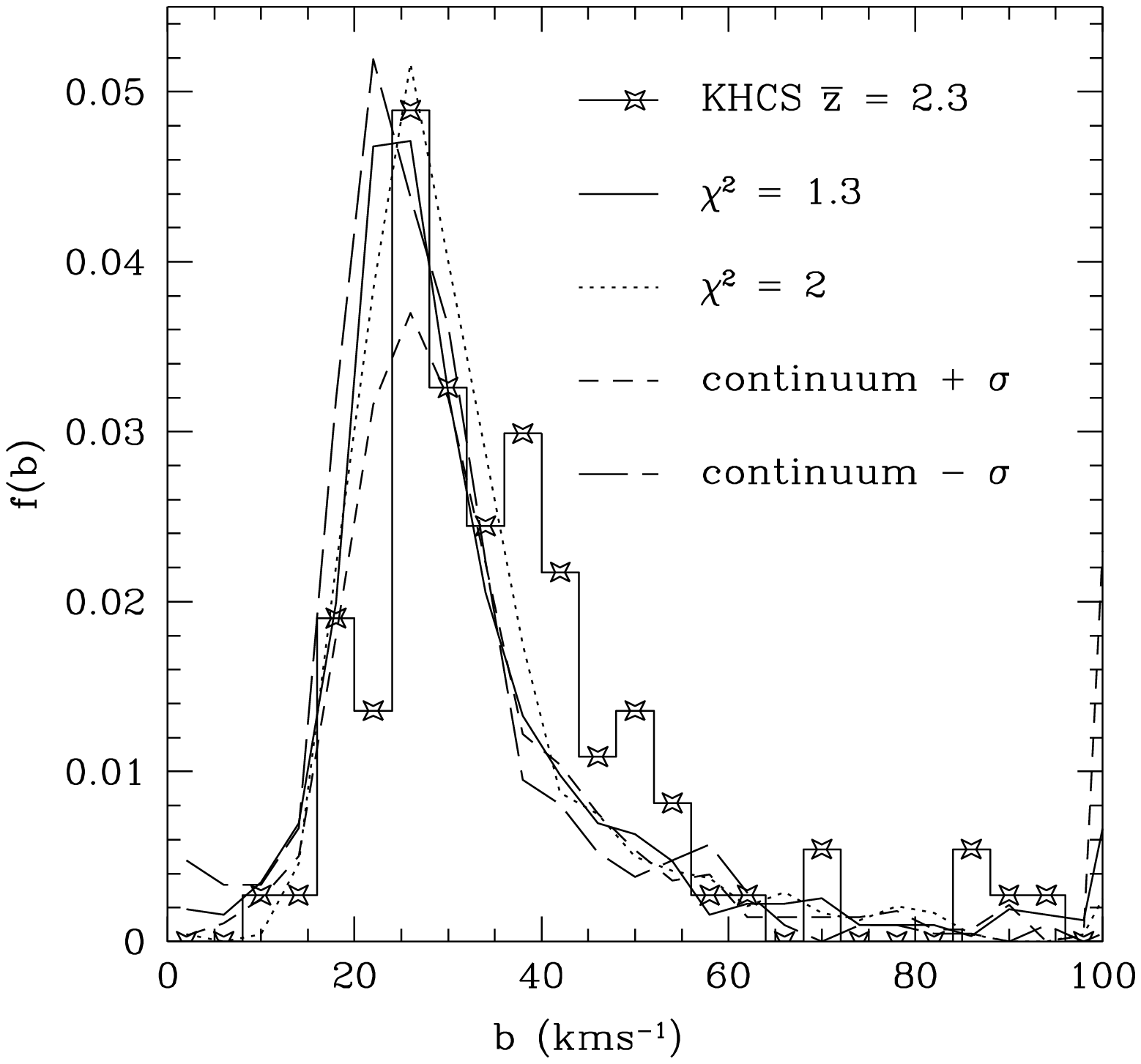}}
\caption{$b$-parameter distributions corresponding to the column
densities shown in figure~\ref{fig:Nobs} for lines with $N_\h >
10^{13}$ cm$^{-2}$. The histogram shows the corresponding observational
results of KHCS. Lines broader than 100 km s$^{-1}$ are arbitrarily set
to 100 km s$^{-1}$.}
\label{fig:bobs}
\end{figure}
The VP-fitting process is sensitive to the quality of the observational
data, the continuum level chosen, and the strictness of the fit
demanded, as we will quantify here. In figure~\ref{fig:Nobs} we show
the DDF of the high resolution A-11-64 simulation at z=2, calculated
using varying VP-fitting parameters. The solid line shows our previous
results obtained using the default reduced $\chi^2$ requirement,
$\chi^2 < 1.3$, and choosing a continuum fit as described
previously. The dotted lines show the results of relaxing the $\chi^2$
requirement to $\chi^2 < 2$, thereby simulating observations with lower
signal to noise. This clearly reduces the number of very weak lines,
$N_{\h} < 10^{12.5}$ cm$^2$, found, but otherwise leaves the DDF
unchanged. The long-dashed line shows the results when the continuum
level found by the procedure outlined above is then lowered by just
2\%, this being the noise level set for these spectra (once again
requiring $\chi^2 < 1.3$). Here we see a similar but even less
noticeable effect. Finally the short-dashed line shows the result of
raising the continuum level by the same amount. We see now that the
column density distribution is slightly raised down to $N_{\h} <
10^{13.5}$ cm$^2$. Beyond this column density it seems that the
distribution found is insensitive to the chosen parameters of the VP
fitting process.

Figure~\ref{fig:bobs} shows the $b$-parameter results given the
different VP analyses described above. Encouragingly, the changes are
slight except for the case where the continuum level is chosen higher
which decreases significantly the height of the observed peak and
correspondingly increases the number of lines at large values of $b\ge
100$ km s$^{-1}$.

\subsection{Comparison with UV background scaling}
\begin{figure}
\resizebox{\columnwidth}{!}{\includegraphics{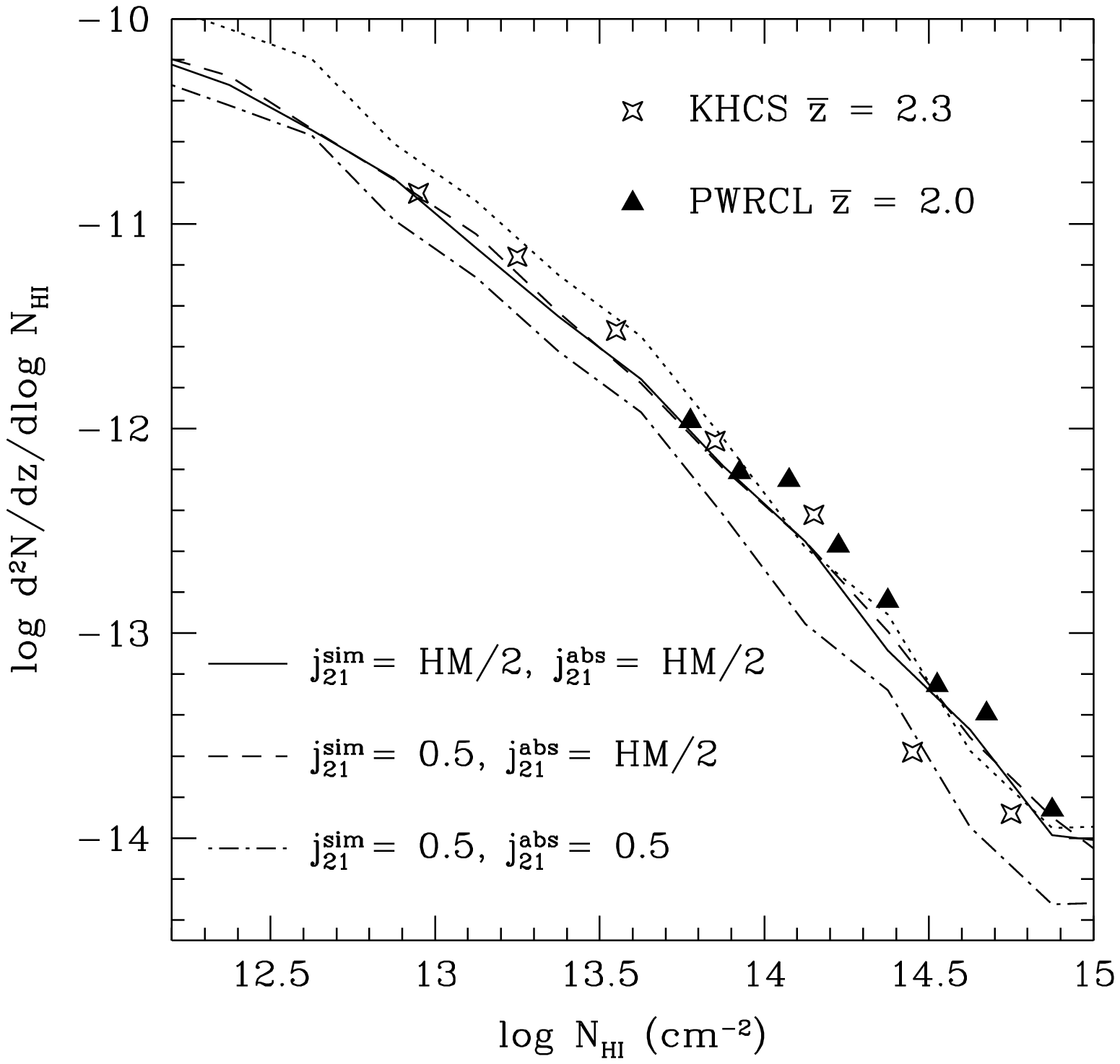}}
\caption{Column density distributions derived from H-11-64-j
simulations at z=2. The solid and dot-dashed lines show results where
the assumed UV backgrounds are given by HM/2 (tabulated UV background
of HM with amplitude divided by 2), and a fixed power-law ($J_{21} =
0.5, \alpha = 1$), respectively both during run-time and for the
calculation of spectra. Results are also shown for the latter
simulation but where an HM/2 UV background is assumed for the
calculation of spectra (dashed line), and where the original results
have been raised by $\log(2.36)$ (dotted line). The latter factor 2.36
is the ratio of the flux at the \lya limit between the $J_{21}=0.5$
power law spectrum and the HM/2 one. Observational results of KHCS and
PWRCL are also plotted as indicated.}
\label{fig:Njscale}
\end{figure}
\begin{figure}
\resizebox{\columnwidth}{!}{\includegraphics{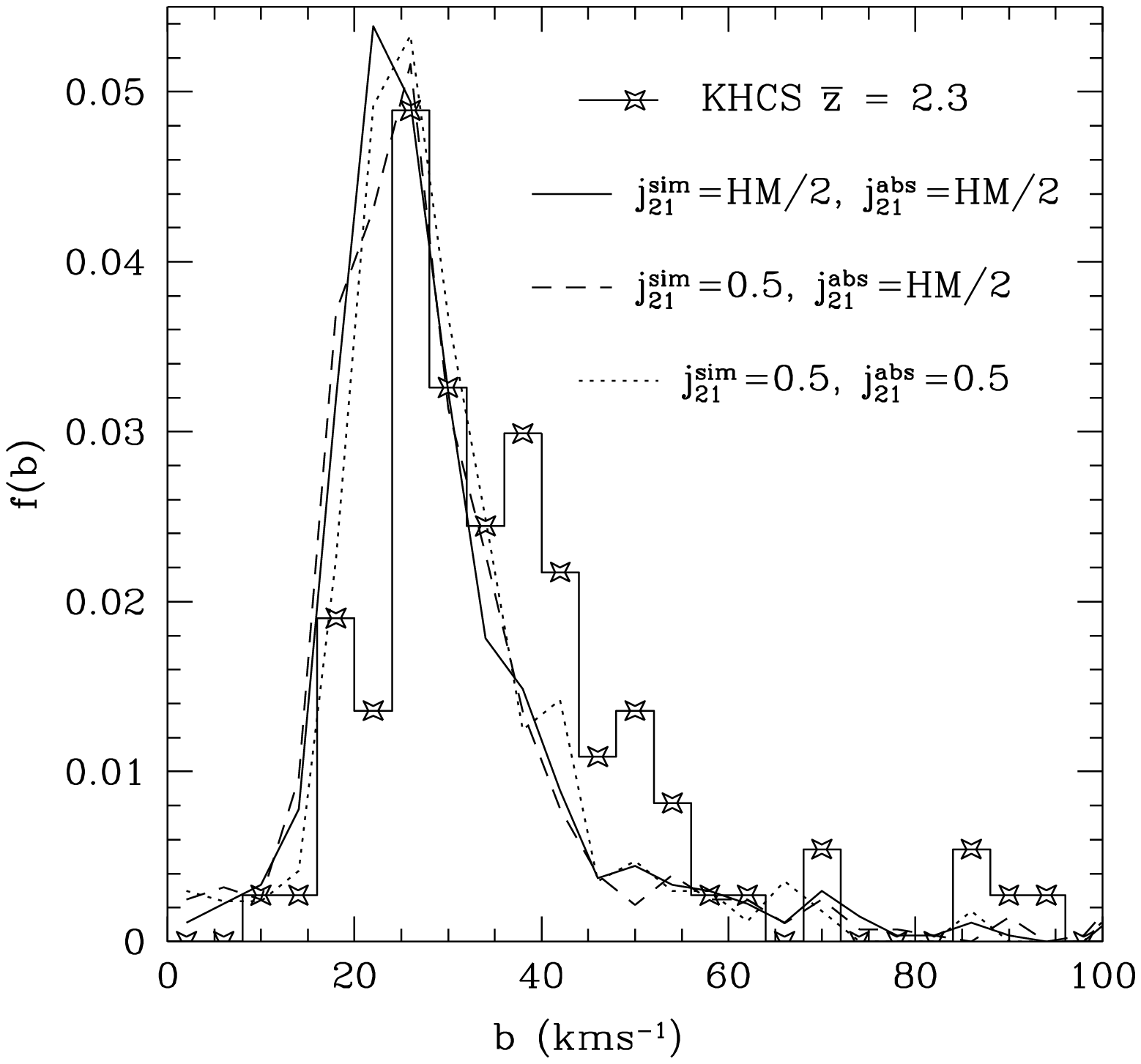}}
\caption{$b$-parameter distributions corresponding to the column
densities shown in figure~\ref{fig:Njscale} for lines with $N_\h > 10^{13}$
cm$^{-2}$. The histogram shows the corresponding observational results
of KHCS.}
\label{fig:bjscale}
\end{figure}
The deduced DDFs and $b$-parameter distributions depend on the
amplitude of the assumed ionising background which is not well known,
and on the baryon fraction which is equally uncertain. The shape of the
radiation spectrum also determines the amplitude and slope of
temperature-density relation obeyed by the cool IGM (see
equation~(\ref{eq:T0}) for the case of a power law radiation spectrum)
and hence has some influence on the neutral fraction of gas at a given
density. This potentially introduces a large parameter space to search
through with numerical simulations. Fortunately, the resulting optical
depth only depends on a particular combination of these
parameters. Indeed, in ionization equilibrium the number of photo
ionizations per unit time balances the number of recombinations, hence
$J_{21}\rho\propto \alpha_\hp \rho^2$. The recombination coefficient
scales with temperature as $\alpha_\hp \propto T^{-0.7}$, so in the low
density phase where $\rho\propto T^{1.7}$ we find $\tau\propto
\Omega_B^{2-0.7/1.7}/J_{21}\propto \Omega_B^{1.59}/J_{21}$ (\eg Hui \&
Gnedin (1997) and Appendix~\ref{App:IGM} for details). The exponent of
$\Omega_B$ depends weakly on the shape of the ionising spectrum. We now
show that simulations can be run with one set of values of $\Omega_B$
and $J_{21}$ and later scaled to good accuracy to another set.
Figure~\ref{fig:Njscale} compares the DDFs for simulations H-11-64 and
H-11-64-j. The first of these was run with the Haardt \& Madau (1996)
ionising background (with amplitude divided by 2, indicated by the
lines labelled \lq HM/2\rq~) and the second with an imposed power law
spectrum of ionising photons with constant amplitude $J_{21}=0.5$ and
slope $\alpha=1$. The DDFs of these two runs are significantly
different, with the H-11-64 run producing more lines at all column
densities. However, if we assume {\em in the analysis phase} that the
spectrum is HM/2 for the H-11-64-j run than we obtain an almost
identical DDF as for the original H-11-64 run. This kind of scaling
also works well for the associated $b$-parameter distributions, shown
in figure~\ref{fig:bjscale}. Note that the run with power law ionising
sources produces a $b$-parameter distribution very similar to the
original HM/2 run, even without post-processing scaling.

\section{Comparison with observations}
\label{sect:observations}
\begin{figure}
\resizebox{\columnwidth}{!}{\includegraphics{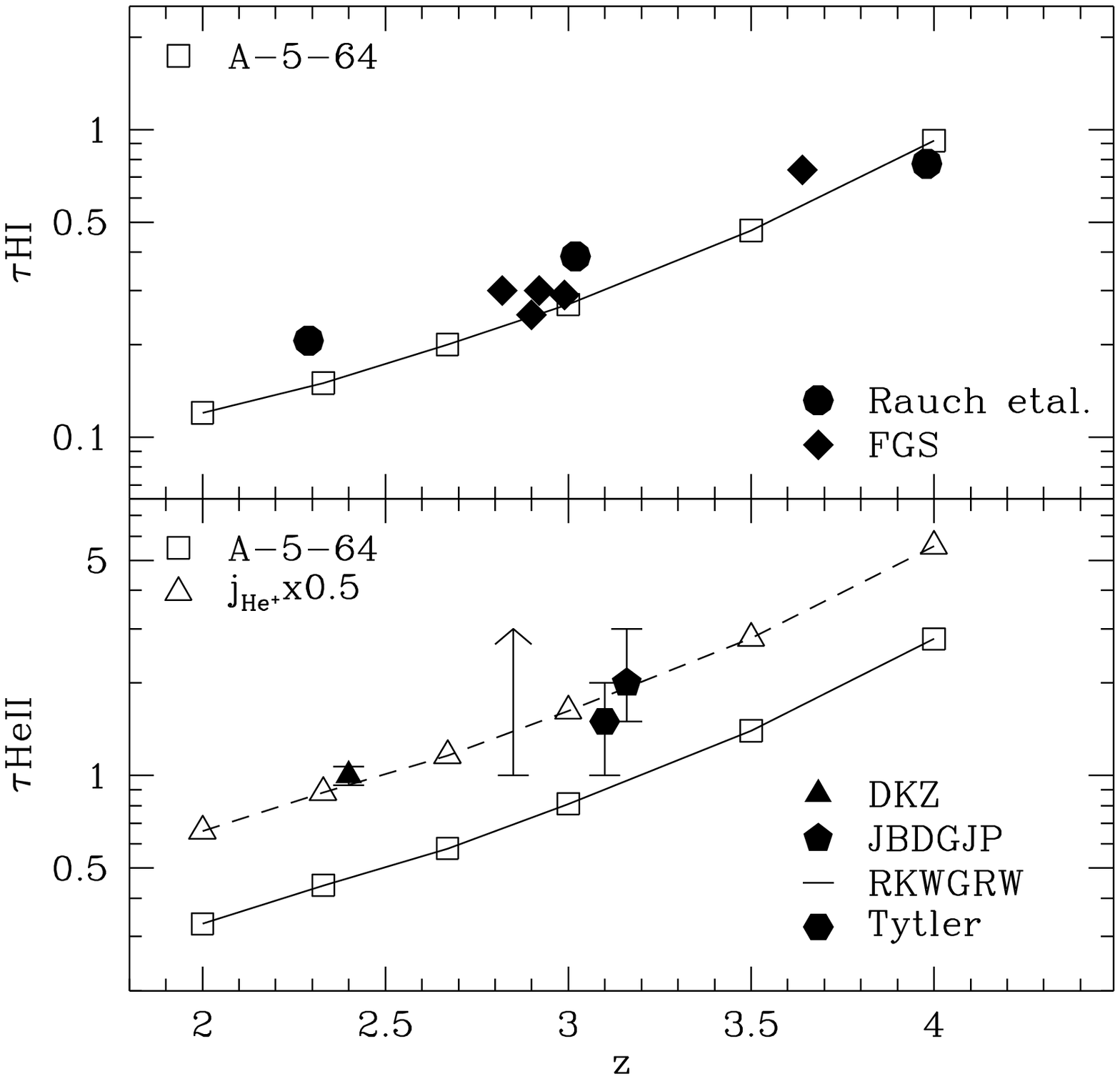}}
\caption{Mean optical depth $\bar\tau$ as a function of redshift $z$
for our highest resolution run A-5-64 (open squares connect by solid
line). The top panel refers to hydrogen, where we have plotted the
observed data points presented in Rauch \etal (1997) and those
determined by Fardal \etal (1998, denoted FGS) using combined
line-lists from authors using high-resolution spectra. The bottom panel
refers to singly ionised helium. The observational data points shown
are those of Davidsen \etal (1996, DKZ), Jakobsen (1997, BDGJP),
Reimers \etal (1998, RKWGRW) and Tytler \etal (1998, as plotted in
figure~1 of Fardal \etal 1998). The dashed line connecting triangles in
the lower panel shows the effect of increasing the
$\Gamma_\h/\Gamma_\hep$ break in the Haardt \& Madau UV background by a
factor 2.}
\label{fig:TauObs}
\end{figure}
\begin{figure}
\resizebox{\columnwidth}{!}{\includegraphics{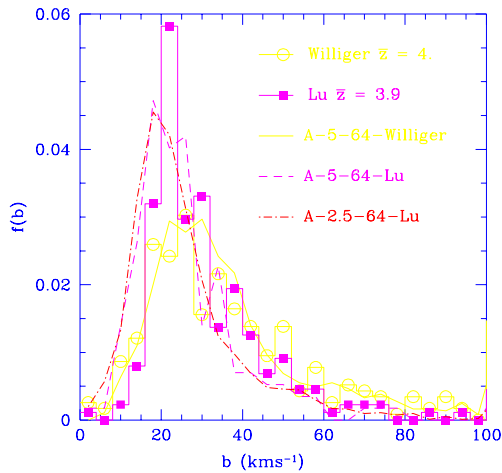}}
\caption{$b$-parameter distribution for lines with $N_\h > 10^{13}$
cm$^{-2}$ at $z=4$ for A-5-64 and A-2.5-64 compared to data from
Williger \etal (1994) and Lu \etal (1996). These two data sets were
taken with different resolution and so we analysed the simulated
spectra accordingly: A-5-64-Williger is analysed assuming a FWHM of 12
km s$^{-1}$ and SNR of 20 in the VPFIT procedure, A-5-64-Lu and
A-2.5-64-Lu with FWHM of 8 km s$^{-1}$ and SNR of 60.}
\label{fig:bfigz4}
\end{figure}
\begin{figure}
\setlength{\unitlength}{1cm}
\centering
\begin{picture}(10,11)
\put(-1., -4.0){\includegraphics{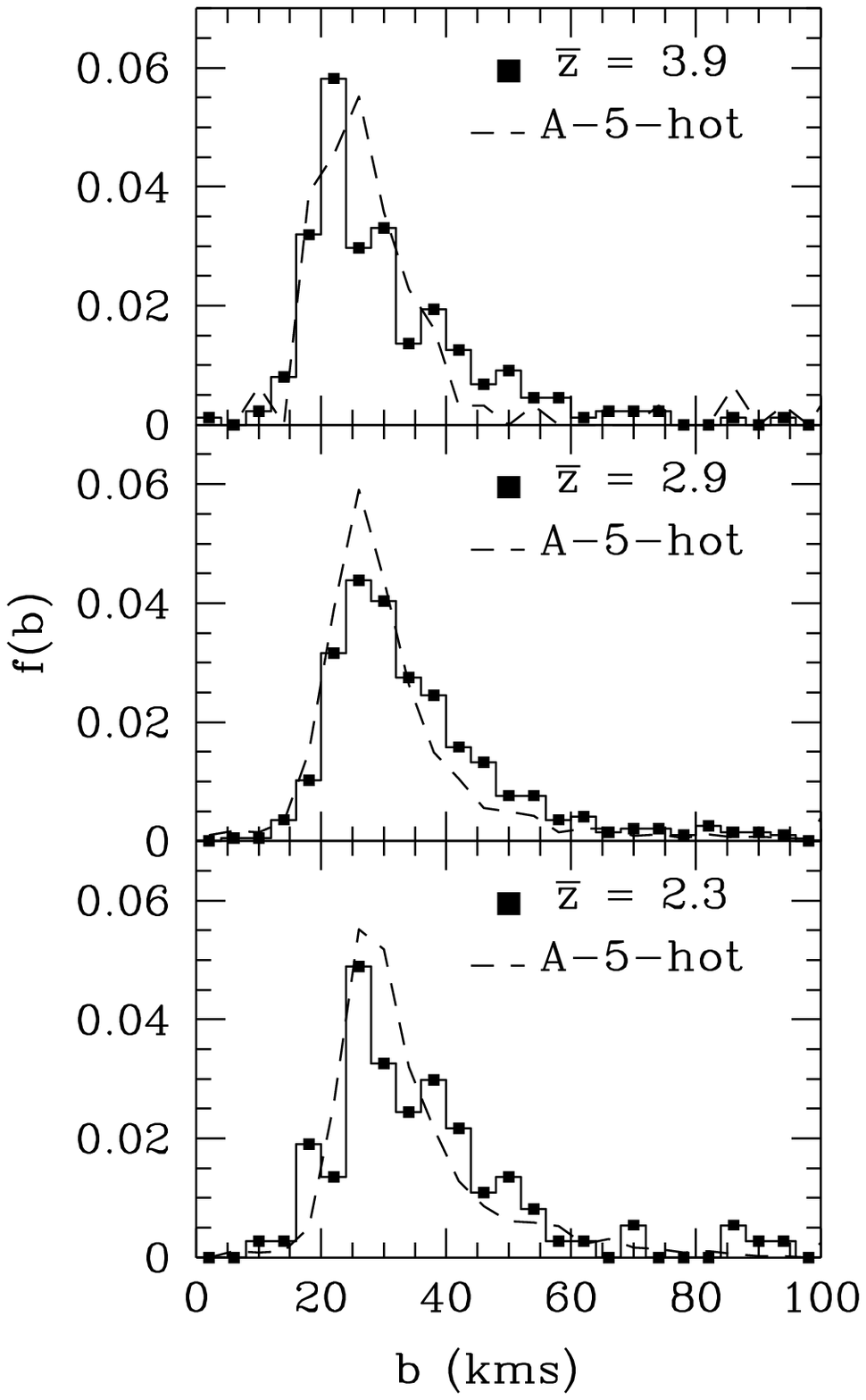}}
\end{picture}
\caption{$b$-parameter distribution for lines with $N_\h > 10^{13}$
cm$^{-2}$ for A-5-64, after increasing the temperature of all gas
arbitrarily by a factor of two, for redshifts 4, 3 and 2 (top to bottom
respectively). The VPFIT parameters in the analysis were chosen to
allow comparison with the data of Lu \etal (1996, $\bar z=3.9$), Hu
\etal (1995, $\bar z=2.9$) and Kim \etal (1997, $\bar z=2.3$), which
are superposed for comparison.}
\label{fig:bfig7}
\end{figure}
Given the level of convergence of simulated quantities, as discussed in
the previous section, we now turn our attention to a detailed
comparison of simulations with observations at redshifts $\ge 2$.

A comparison of the effective optical depths for \H and \Hep, computed
from the A-5-64 run at z=2, with observational data is given in
figure~\ref{fig:TauObs}. For the evolution of $\tau_\h$ we plot the
recent results of Fardal \etal (1998), who used a combined line-list
from a variety of authors for which only those originating from studies
carried out using the HIRES spectrograph on the Keck were included
here. We also show the effective optical depths observed by Sargent \&
Barlow (published in Rauch \etal 1997). It is clear that the simulated
$\tau_\h$ results fit well, as might be expected from the good fit of
the \H DDF results. Rauch \etal (1997) used the good agreement between
simulated and observed optical depths to set a lower limit on the
baryon fraction, using a lower limit on the ionization flux deduced
from the observed quasar luminosity function. Since we confirm their
simulation results for standard CDM models, we also confirm their lower
limit, $\Omega_B h^2 \gtsima 0.017$.

For $\tau_\hep$ we have used the recent \Hep Gunn-Peterson effect
detections collected by Fardal \etal (1998) (see references
therein). Although the available $\tau_\hep$ observations are limited,
the results of Davidsen \etal (1996) do supply a strong constraint at
$\bar z_{\rm abs} = 2.4$. In this case our result that the $\tau_\hep$
value inferred from our simulations does indeed change significantly
when simulating at higher resolution, means that this constraint is not
matched by our simulations which predict a $\tau_\hep$ lying about a
factor 2 below the observed value. Unlike the conclusions of Croft
\etal (1997) we consequently find it is impossible to match both the
$\tau_\h$ and the $\tau_\hep$ observational constraints by a applying a
single renormalisation to the Haardt \& Madau (1996) UV background
spectrum, and instead require a 'softer' spectral shape. Since
$\tau_\hep$ scales as $\tau_\hep/\tau_\h \propto \Gamma_\h/\Gamma_\hep
\propto J_\h/J_\hep$, where $J_\h/J_\hep$ is the ratio of the ionising
flux at the respective \H and \Hep limit frequencies, we can increase
the helium optical depth by scaling the helium ionization rate keeping
the hydrogen photo-ionization rate constant. Increasing the ratio
$J_\h/J_\hep$ by a factor two at all observable redshifts from $\sim 7$
to $\sim 14$ then provides a good fit to both the $\h$ and $\hep$
optical depths (figure~\ref{fig:TauObs}).

This softened UV background may yet prove to be consistent with the UV
background inferred from quasars alone, since recent estimates of the
\lq average\rq~ intrinsic quasar spectral index have yielded softer
values $\sim 1.8$ (Zheng \etal 1997), as opposed to the value $1.5$
originally assumed in Haardt \& Madau (1996). Nevertheless, since the
values of the UV background at the ionising edges are strongly affected
by the self-absorption by \lya forest clouds, we must wait for more
detailed models taking these effects into account before drawing any
strong conclusions. The values for $\Gamma_\h$, and $\Gamma_\hep$
depend somewhat on the effective slope of the spectrum at frequencies
immediately higher than their ionising frequencies, however this
dependence is weak, so achieving the required factor two change in the
ratio of photo-ionising coefficients by changing the background
spectrum in this way is unlikely.

Previously we have have shown that our SCDM simulations using the
Haardt \& Madau (1996) spectrum (with amplitude divided by 2) reproduce
the observed DDFs quite well, both at $z=2$ (figure \ref{fig:Nnumer2})
and $z=3$ (figure \ref{fig:Nnumer3}). The only discrepancy may be that
the simulations produce a slightly lower DDF than seen at $z=2$ by
Petitjean \etal (1992). The difference is slight and in any case the
observations are as yet not discriminating enough at this redshift to
really suggest that this is a problem, as the newer results of Kim
\etal (1997) at $z=2.3$ are somewhat discrepant with the older data.
However it is not clear at present to what extent the DDF discriminates
different plausible cosmological models and consequently it is not yet
possible to use the relatively good agreement between observed and
simulated DDFs as an argument in favour of our assumed cosmology.

The simulated and observed $b$-parameter distributions at $z=4$ are
compared in figure~\ref{fig:bfigz4}. We have analysed the simulated
spectra using a VPFIT set-up that mimics the signal to noise ratio and
spectral resolution of the data from Williger \etal (1994, simulations
shown as A-5-64-Williger) and Lu \etal (1996, simulations labelled
-Lu). For the latter data we compare A-5-64 with the higher resolution
A-2.5-64 run. First note the excellent agreement between the different
resolution runs, suggesting that the $b$-parameter distribution in the
5Mpc box has very nearly converged. We noted earlier that the \Hep
optical depth of these two runs is very similar as well, which
increases our confidence in the reliability of results drawn from the
5Mpc box run, as far as numerical artifacts such as resolution are
concerned. Next note the importance of the wavelength resolution in the
analysis stage, by comparing the A-5-64 run analysed with two different
VPFIT set-ups.  The lower spectral resolution of A-5-64-Williger versus
A-5-64-Lu reduces dramatically the peak in the $b$-distribution at
$\sim$ 25 km s$^{-1}$, correspondingly increasing the number of broader
lines. This trend is also shown by the data. The $b$-parameter
distribution of A-5-64-Williger provides an excellent fit to the data
from Williger \etal (1994), both at low and high $b$ values. At higher
resolution, when we compare the A-5-64-Lu curve with the Lu \etal
(1996) data, the agreement is still very good. There is however a hint
that there are too many narrow lines in the simulated spectra.

This difference between simulated and observed distributions at small
$b$-parameters is even clearer at lower $z$ (see
figure~\ref{fig:bnumer2} for $z=2$ and figure~\ref{fig:bnumer3} for
$z=3$, comparing data with the resolved A-5-64 run): the simulated
$b$-parameter distributions peak at lower values of $b\sim 20$ km
s$^{-1}$ whereas the observed ones peak at $b\sim 30$ km s$^{-1}$. A
fair comparison between observations and simulations is hampered by the
sensitivity of the $b$-parameters to the continuum fitting
procedure. Note that the high resolution simulations have a box size of
only $\sim 10{\mbox \AA}$, and this might introduce large continuum
errors. Also, the observed $b$-parameter distributions have been shown
to vary slightly from quasar to quasar (Kim \etal 1997), though the
effect is larger for the higher column density systems $N_\h >
10^{13.8}$ cm$^{-2}$. Despite these complicating factors the simplest
interpretation of our results is that our simulated IGM has
temperatures that are slightly lower by a factor $\sim 2$ than those
existing in the actual absorbing IGM. Indeed, if we increase
arbitrarily the temperature of the simulated gas by a factor of two
(dashed lines in figure~\ref{fig:bfig7}) then we find excellent
agreement between the simulated and observed $b$-distributions at all
redshifts 2, 3 and 4, in accord with the findings of Haehnelt \&
Steinmetz (1998).

What could give rise to such a hotter IGM? As discussed in Section 2.1,
the temperature of the IGM at redshifts $z=4\rightarrow 2$ depends on
the epoch of reionization. In the Haardt \& Madau scenario, \H
reionizes at $z\sim 6$ and \Hep at $z\sim 4.7$ (dashed line in
figure~1b), and the reionization is due to the increase in the combined
QSO luminosity at those high redshifts. However, at present there are
no known QSOs at $z>5$, hence the assumed increase in QSO flux depends
mostly on the extrapolation of the evolution of the QSO luminosity
function to higher $z$. If \Hep reionization would be delayed, then
non-equilibrium effects would give a substantial increase in IGM
temperature, even at redshifts of $z\sim 2$. Consequently,
uncertainties in the QSO flux at redshifts $\ge 5$ alone would appear
to give sufficient leverage for the required increase in the IGM
temperature. It is not clear whether such changes in reionization
history alone could conspire to give the correct increase in the mean
IGM temperature over the whole of the redshift interval $4\rightarrow
2$. Note that any increase in temperature necessary to provide
$b$-distributions that fit observations, would in effect scale both the
DDF and $\tau_\h, \tau_\hep$ results further; increasing $T$ by a
factor of 2 requires increasing $\Omega_B$ by a factor $\sim 2^{0.7}$
to keep the same level of absorption.

There exist some further numerical possibilities that may result in a
simulated IGM that appears slightly cool. Firstly, larger waves not
included in our highest resolution run of box-size 5.5Mpc could
dynamically heat the medium to a greater extent than found in our
simulations. The lack of dependence of the $b$-distribution to the
different box-sizes examined here indicates that this is probably not
an important effect. Secondly, baryons collapsed in small dark matter
halos that have formed at high-$z$ could possibly be \lq evaporated\rq~
by a large temperature boost at reionization such as found in the
non-equilibrium models. Our smoother temperature increase during
reionization may then allow more low-circular velocity halos to
capture gas than appropriate. To reiterate then, we find that it is
likely that a temperature boost within our simulations is likely to be
necessary in order to produce absorption lines that fit the
observations at $z=2$, 3 and 4. The necessary temperature increase may
well result from a consistent treatment of the thermal history as
computed by Haardt \& Madau (1996) but including non-ionization
equilibrium effects, though our estimates show that higher temperatures
due to an even later reionization epoch, may eventually be
necessary. Another plausible candidate for extra energy input into the
gas is feedback from star formation. Certainly it is likely that in
future observations of the Doppler parameter distribution of the \lya
forest could become an excellent tool in providing constraints on the
thermal history of the universe beyond $z=2$.

\section{Summary and conclusions}
\label{sect:summary}
We have presented a new simulation tool, \apm, designed to study
numerically the formation of structures responsible for the
\lya-forest. This code is very fast and treats the low density IGM
relatively accurately, allowing increased resolution at little extra
simulation time. The IGM is allowed to interact with a time-dependent
but uniform background of ionising photons assumed to come from
quasars, using the rates suggested by Haardt \& Madau (1996). This
background heats the low density gas and changes the form of the
cooling function at higher densities. The distribution of the gas in
the density-temperature plane can be understood from the relative
importance of cooling and heating processes, and from the comparison of
the appropriate cooling time scale with the Hubble time.

We performed extensive comparisons of the new code with the \hydra code
of Couchman \etal (1995), which was adapted to study this problem as
well. The agreement between the two codes is excellent for a wide
variety of statistics. The distribution of gas in the $(\rho,T)$ plane
is very similar and various statistics on the distribution of halos
agree very well. The amount of gas which is able to cool in collapsed
halos is similar in the two codes, showing that coding details are not
very important in determining this fraction. We are currently analysing
several large \hydra simulations performed on the T3D computer to
understand in more detail how resolution affects \lya statistics (the
VIRGO consortium, in preparation). Both \apm and \hydra are based on
the Lagrangian SPH method, which has high resolution in high density
regions. However, since many lines form in {\em low} density regions
where SPH suffers from low resolution, it would still be very valuable
to compare in detail with some of the Eulerian codes used by other
groups.

We also compared our new code with published results from \tree
(Hernquist \etal 1996) for simulations started from their initial
conditions and confirm their findings. We have also analysed
independently our simulated spectra from these runs using a different
implementation of automated Voigt profile fitting (VPFIT, Carswell
\etal 1987). The deduced line statistics in terms of column density and
$b$-parameter distributions agree well with their published values,
showing that Voigt profile fitting gives reproducible results.

We have then used \apm to study the effects of lack of numerical
resolution on quantities deduced from simulated spectra based on Voigt
profile fitting. The mean effective hydrogen optical depth is converged
in our medium resolution simulation and so are the derived column
density distributions (DDFs). The latter also are in good agreement
with DDFs deduced from observations, for our assumed background flux
and baryon fraction. However, the relative amounts of cool gas are
rather different when comparing the A-22-64 with the A-11-64 run, which
has eight times better mass resolution, and there are still noticeable
differences with our highest A-5-64 run (which has another factor of
eight better mass resolution), due to lack of numerical
resolution. With increasing resolution, we find that the optical depth
decreases, especially for \Hep, and that the number of lines with
small $b$-parameter increases. However, from a comparison of the A-5-64
run with an even higher resolution simulation, A-2.5-64, we find that
the A-5-64 box is already very close to convergence and we are
relatively confident that we can draw reliable conclusions from this
simulation. We found that the deduced $b$-parameter distributions are
sensitive to the assumed continuum level, a problem which should also
influence observations to some extent. The DDFs, on the other hand, are
not very sensitive to the exact VP-fitting procedure.

Some previously published results on the \Hep forest are unreliable due
to lack of numerical resolution. For example at $z=4$, the mean
effective \Hep optical depths are 4.54, 3.52, 2.78 and 2.63 for runs
A-22-64, A-11-64, A-5-64 and A-2.5-64, respectively. This shows that
the required resolution to get the mean optical depth correct is very
high. We interpreted the dependence on resolution as being due to the
formation of progressively smaller halos being resolved with better
resolution. Low density gas falls into these halos and hence the
optical depth decreases. The good agreement between the 5.5Mpc and the
2.5Mpc box increases our confidence that these higher resolution runs
have effectively converged.

Turning to a comparison of our highest resolution simulation with
observations, we come to the following conclusions.

\begin{itemize}
\item There is excellent agreement between the observed and simulated \lya
column density distributions at $z=2$ and $3$, provided we divide the
ionising background intensity advocated by Haardt \& Madau (1996) by
two, for our assumed baryon fraction of $\Omega_B
h^2=0.0125$. Alternatively, for the intensity of the ionising
background as computed by Haardt \& Madau, we require a higher baryon
fraction (Rauch \etal 1997)
\begin{equation}
\Omega_B h^2 \gtsima 0.017\,\,\,\,{\mbox{\rm (from DDFs)}}\,.
\end{equation}

\item The simulated $b$-parameter distributions peak at lower
$b$-values than the observed ones for $z=4$, 3 and 2, suggesting that
the simulated IGM temperature in our simulations is too low. We argued
that uncertainties in reionization history, combined with
non-equilibrium effects and feedback from star formation, might be
sufficient to increase the temperature by a factor $\sim 2$, which
would bring the simulated distributions into excellent agreement with
the observed ones. However, this would increase the required $\Omega_B$
even more, since increasing the temperature would decrease the amount
of absorption, giving the higher $\Omega_B$ limit

\begin{equation}
\Omega_B h^2 \gtsima 0.028\,\,\,\,{\mbox{\rm (from $b$-parameter distribution)}}\,.
\end{equation}

\item The \Hep optical depth corresponding to our best fit \H optical
depth is lower than observed values, suggesting that the Haardt \&
Madau ionization spectrum may be too hard. The more recent analysis by
Zheng \etal (1997) of observed quasar spectra lead to a similar
conclusion. Fitting both \H and \Hep optical depths requires a spectral
break
\begin{equation}
{J_\h\over J_\hep} \approx 14\,.
\end{equation}

\end{itemize}

Overall we find that the level of agreement between simulations of the
\lya forest in a scale-invariant, CDM universe and observations, is
still impressive. More detailed comparisons between simulations and
observations will allow us to study the thermal history of the universe
at even higher redshifts.

\section*{Acknowledgements}
TT acknowledges partial financial support from an EC grant under
contract CT941463 at Oxford University. APBL thanks PPARC for the award
of a research studentship and GPE thanks PPARC for the award of a
senior fellowship. We thank H. Couchman for making his P3M code
available as the basis for \apm. We are indebted to R. Carswell for
help with VPFIT and helpful discussion, to T. Quinn for help with TIPSY
and to R. Croft for given us the initial conditions from the \tree
simulations. We thank M. Haehnelt for many stimulating discussions and
suggestions on the manuscript.

{}

\appendix
\section{Mathematical description}
In the first part of this Appendix we give details of the equations
describing the growth of structure. The second part gives the explicit
equations for computing SPH quantities in the \apm implementation as
well as a detailed description of the way we determine SPH
neighbours. The third part of this Appendix describes the \hydra low
density correction. In the final part of this Appendix we give the explicit
expressions used to compute simulated spectra.
\label{sect:maths}
\subsection{Physical model}
\label{sect:eqs}
In the Newtonian approximation valid on the scales under consideration,
the evolution of structures is governed by the following set of
Lagrangian equations:
\begin{eqnarray}
{d\hat\rho\over dt} &=& -\hat\rho \nabla\bfx \\
{d\dot\bfx\over dt} + 2 {a\over\dot a} \dot\bfx&=& -{1\over a^3} \grad\psi 
- {1\over a^2} {\grad p\over\rho}\label{eq:Euler}\\
{d u\over dt} + 3 {\dot a\over a} {p\over \rho} &=& -{p\over\rho}
\nabla\dot\bfx + \left({1-Y\over m_H}\right)^2\,\rho ({\cal H}-{\cal
C})\label{eq:thermal} \\
\nabla^2\psi &=& 4\pi G (\hat\rho_{\rm T} -\langle\hat\rho_{\rm
T}\rangle)\label{eq:Poisson}.
\end{eqnarray}
Here, $\hat\rho\equiv a^3\rho$ is the comoving gas density, $Y$ is the
helium abundance by mass, ($1-Y$) is the hydrogen abundance,
$\bfx=\bfr/a$ are comoving coordinates, $\bfv_p\equiv a\dot\bfx$ is the
peculiar velocity, $a(t)=(1+z)^{-1}$ is the scale factor, $t$ denotes
time, $z$ is the redshift and $\nabla\equiv\partial/\partial\bfx$. The
pressure is $p=(\gamma-1)\rho u$, where $u$ is the thermal energy per
unit mass and $\gamma=5/3$ for a mono-atomic gas. The density
$\hat\rho_{\rm T}$ entering in the Poisson equation (\ref{eq:Poisson})
is the sum of the dark matter and gas density; $G$ is the gravitational
constant and $m_H$ is the proton mass. The dark matter evolves
according to the Euler equation (\ref{eq:Euler}) with $p=0$. These
equations neglect feedback from stars and AGN. Defining
\begin{eqnarray}
K &=& \int {\hat\rho\over 2} \bfv_p^2\, d^3\bfx\\
U &=& \int \hat\rho u \,d^3\bfx\\
W &=& \int \hat\rho\psi\,d^3\bfx\\
L &=& \int \hat\rho^2 ({\cal H} -{\cal C}) / a^3\, d^3\bfx
\end{eqnarray}
one can write the Layzer-Irvine cosmic energy equation as (\eg Peebles
1980, \S 24)
\begin{eqnarray}
\Delta I &=& \int_{a_i}^a (W + (5-3\gamma) U + a^2 L/\dot a) da \nonumber\\
&+& a \int_{a_i}^a (K+(3\gamma-4)U-a L /\dot a ) da \nonumber\\
&-& (a-a_i) a_i(K_i+U_i+W_i)\nonumber\\
&=& 0,
\label{eq:Layzer}
\end{eqnarray}
where the index $i$ means at the initial expansion factor $a_i$.

The functions ${\cal H}(\rho, u)$ and ${\cal C}(u)$ in
equation~(\ref{eq:thermal}) describe the heating of the medium by
photo-ionization and cooling through collisions and interaction with
the CMB, respectively. In our simulations we use the evolution of the
photo-ionising background as computed by Haardt \& Madau
(1996). Detailed expressions for the fits to the temperature dependence
for all included processes are given in Appendix~\ref{sect:cooling}.

\subsection{\apm implementation}
\label{sect:SPH}
The explicit expressions to compute the SPH quantities of particle $i$
are
\begin{eqnarray}
\hat\rho(i) \bbbb &=&\bbbb\sum_j {\cal W}_{ij}
\label{eq:SPHdens} \\
h \nabla_\bfr \bfv (i)\bbbb &=&\bbbb
-{h(i)\over\rho(i)}\,\sum_j (\bfr(i)-\bfr(j))\cdot(\bfv(i)-\bfv(j))
d{\cal W}_{ij}\label{eq:SPHdivv}\nonumber\\
\\
{\grad p\over\rho}(i) \bbbb &=&\bbbb \sum_j \left({s(i)^2\over
\hat\rho(i)}+{s(j)^2\over \hat\rho(j)}\right) (\bfx(i)-\bfx(j))
\,d{\cal W}_{ij}\label{eq:SPHgradp} \\
{p\over\rho}\nabla\dot\bfx (i) \bbbb &=&\bbbb {s(i)^2\over\hat\rho(i)}\, \sum_j
(\bfx(i)-\bfx(j))\cdot(\dot\bfx(i)-\dot\bfx(j))\,d{\cal W}_{ij}
\label{eq:SPHdu}\,.\nonumber\\
\end{eqnarray}
Here, ${\cal W}_{ij} = m W(q_{ij})/h_{ij}^3$ is the normalised SPH
kernel and $d{\cal W}_{ij} = m \partial W(q_{ij})/\partial
q_{ij}/q_{ij} /h_{ij}^5$ its derivative; $m$ is the SPH particle mass
which is the same for all SPH particles. For $W$ we use the M4 spline
(Monaghan 1992) given by
\begin{eqnarray}
W(q) &=& {1\over 4\pi} (4-6q^2+3q^3) \,\,\,{\rm if}\,q\le 1\nonumber\\
     &=& {1\over 4\pi} (2-q)^3 \,\,\,{\rm if}\,1\le q\le 2\nonumber\\
     &=& 0 \,\,\,{\rm else.}
\label{eq:kernel}
\end{eqnarray}
For $q\le 2/3$ we take $(1/q) dW(q)/dq=1/\pi q$ in the calculation of
accelerations to avoid the occurrence of dense knots of SPH particles
within a gravitational smoothing length. We have defined
\begin{eqnarray}
q_{ij} &=& {|\bfx(i)-\bfx(j)|\over \hat h_{ij}} \\
\hat h_{ij} &=& {1\over 2} (\hat h(i) + \hat h(j))\\
s(i)^2 &=& {c(i)^2\over\gamma} + \Pi (i),
\end{eqnarray}
where the artificial viscosity is the sum of a bulk and a von-Neumann
component:
\begin{equation}
\Pi = -\alpha  h \rho c \nabla_\bfr\bfv + \beta \rho h^2 (\nabla_\bfr\bfv)^2.
\end{equation}
Note that the latter is in physical (as opposed to comoving)
coordinates: $h=a\hat h$ is the physical smoothing length, $c=(\gamma
(\gamma-1) u)^{1/2}$ is the physical sound speed and $\bfv=(\dot
a/a)\bfr+a\dot\bfx$ is the velocity. As is usual, the resolution length
$h$ is taken such that on average 32 particles (\lq neighbours\rq~) are
within $2h(i)$ from particle $i$. For computational efficiency we do
not allow $h$ to drop below 1/2 the gravitational spline softening. We
typically take $\alpha=1$, $\beta=2$. SPH quantities are computed from
the particles' positions and velocities in two passes over all the
neighbours: one pass to compute density and velocity divergence
(equations \ref{eq:SPHdens}-\ref{eq:SPHdivv}) and a second pass to
compute the terms entering in the computation of accelerations and
thermal energy derivative (equations \ref{eq:SPHgradp}-\ref{eq:SPHdu}).

Neighbours (particle $j$ is a neighbour of $i$ if its distance to $i$
is smaller than twice the SPH smoothing length of $i$) are found using
a linked-list (Hockney \& Eastwood 1988, p. 274). A square grid is
placed over the computational volume and the linked-list is used as a
book-keeping tool to find which set of particles resides in a given
cell. Since only nearest neighbour {\em cells} are used to check for
potential particle neighbours during the SPH loop, to find all
neighbours for all particles then requires a cell size $\Delta = 2
h_{\rm max}$, with $h_{\rm max}$ the maximum smoothing radius of any
particle. The usage of such a large cell size $\Delta$ becomes rapidly
prohibitively expensive in evolved systems that have a large dynamic
range in density and hence in $h$. We circumvent this problem by using
a hierarchy of cell sizes: we first loop over all neighbouring cell
pairs using a cell size $\Delta$ but only compute interactions between
particle pairs if at least one of its members has $f h_{\rm max}\le
h(i)$. Typically, we take $f=0.8$. In the next pass, $h_{\rm max}$ of
those particles for which all forces have not been computed yet is now
smaller, and so we can do a new loop with a smaller cell size for the
linked-list until all particle pairs have been processed. Consequently,
interactions between particles in high density regions are computed
efficiently with a small linking cell size yet all potential neighbours
of particles residing in low density regions are still found. For
systems with a small dynamic range, this extra book-keeping leads to a
small increase in CPU-time but it leads to huge time savings for more
clustered systems. In fact, the CPU time per step for the SPH
calculations increases only by a factor of 1.6 from a redshift of 50 to
the final highly clustered redshift of 2 (for a simulation using
$2\times 64^3$ particles in an $L=22$~ Mpc box), whereas the CPU time
for the gravity calculation increases by a factor 3.1 over this range.

\subsection{\hydra low density correction}
\label{sect:densityfix}
\begin{figure}
\resizebox{\columnwidth}{!}{\includegraphics{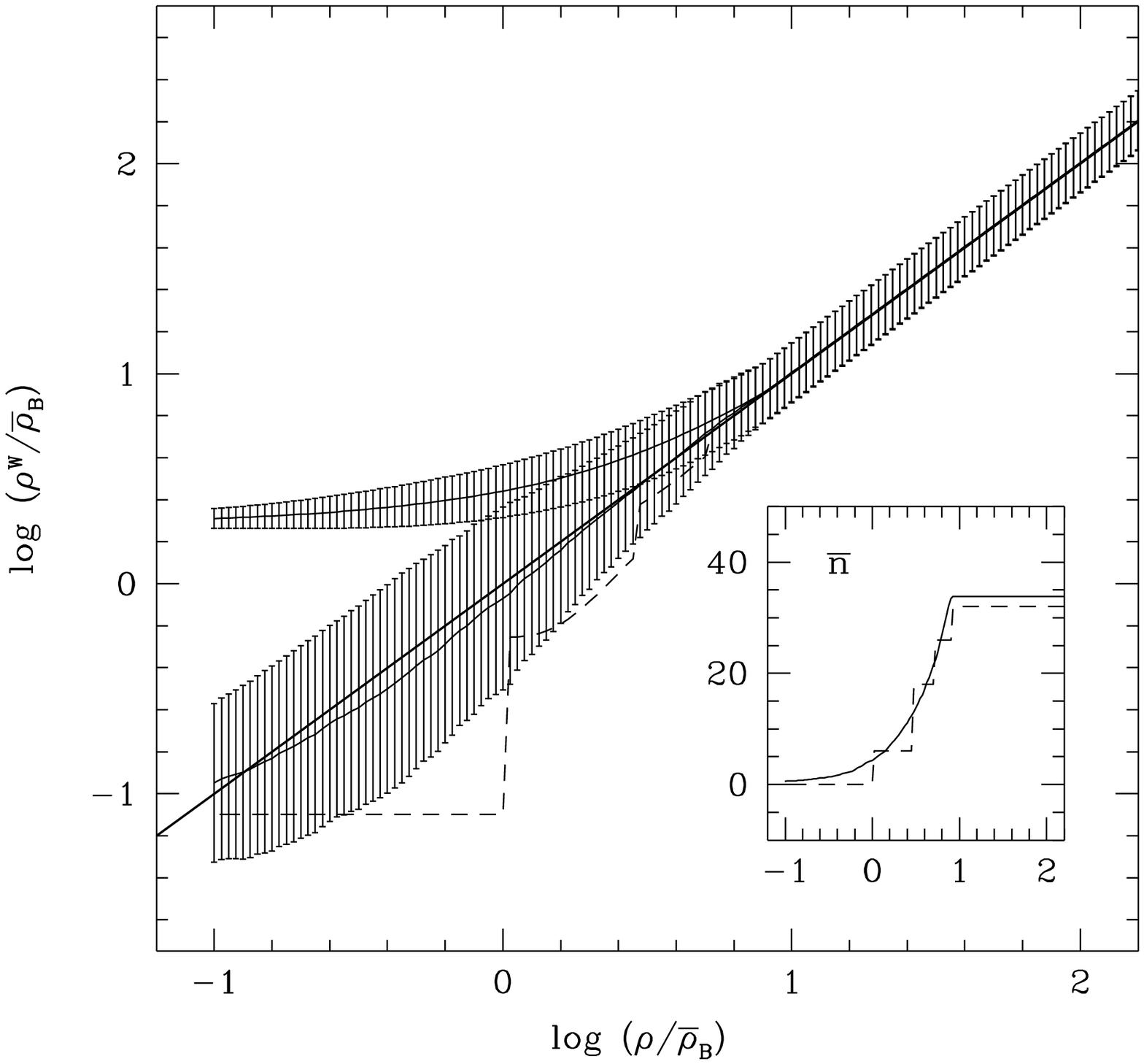}}
\caption{Densities calculated using kernels given by
equations~(\ref{eq:origdens}) (upper shaded region) \&
(\ref{eq:fixdens}) (lower shaded region) of randomly distributed gas
{\em vs} true density, where the SPH neighbour search length, $h <
h_{\rm max}$, so that the the number of neighbours found per particle,
$n$ (inset) can drop to small values at low densities (see text for
details). The dashed lines give the results for $n$, and the density
calculated using equation~(\ref{eq:fixdens}) where gas is distributed evenly
on a grid.}
\label{fig:hydrafix}
\end{figure}

We now describe the correction to the SPH kernel we employed in \hydra.
The effect on the density attributed to a particle when the SPH search
length is restricted is directly dependent on the form of the SPH
kernel, $W$. In fact \hydra uses exactly the same SPH formalism and
kernel as \apm, so using equations~(\ref{eq:SPHdens}) \& (\ref{eq:kernel}) we
may write the density attributed to any particle as
\begin{eqnarray}
\label{eq:origdens}
\hat\rho_i &=& \hat\rho_i^0 + \sum_{j=1,j\neq i}^n {\cal W}_{ij}
\end{eqnarray}
where the particle's `self-density' contribution, $\hat\rho_i^0 = {\cal
W}_{ii}$, is written explicitly ($n$ is the number of neighbours found
within the SPH search length 2$h_i < 2h_{\rm max}$). We illustrate the
results of using this kernel at low densities where $h_i \sim h_{\rm
max}$ and $n$ is small in figure~\ref{fig:hydrafix}. Here the upper
shaded region shows the results of using this kernel to find the
densities of $16^3$ particles randomly filling a box of varying size,
chosen to represent the range of given baryonic over densities
shown. Each particle's search length, $h_i$, was set to the mean
inter-particle separation, where this was less than $h_{\rm
max}$. $h_{\rm max}$ itself was set at half the mean inter-particle
separation at over density unity, as this would be the gravitational
cell-size imposed during a standard \hydra run. With decreasing
$\rho/\bar{\rho}_B \ltsima 10$, the number of neighbours found begins
to drop (inset) due to the search length restriction, and the
calculated particle densities fall slower than do their true densities,
towards a minimum floor when $n=0$. The value for the minimum density
is just the `self-density' contribution specified by the kernel,
$\hat\rho_i^0 = 8/\pi$ in this case.

Since this minimum density is an arbitrary consequence of the kernel
used we are essentially free to reset this minimum self-density term as
we like. We chose the simplest possible modification, calculating
densities according to the following equation for all particles with $n
< $ twenty two
\begin{eqnarray}
\hat\rho_i &=& \frac{{\cal W}_{ii}}{32} + \sum_{j=1; j\neq i}^n
\left({\cal W}_{ij}+{{\cal W}_{ii}\over32}\right) \quad,\quad n < 22  
\label{eq:fixdens}
\end{eqnarray}
while for all particles with $n \ge 22$, the original kernel \ie
equation~(\ref{eq:origdens}) is still applied. Using this compensated
kernel, the self-density level is reduced to a value comparable to the
lowest densities seen in the simulations of \apm, but the kernel itself
is {\em added} to, such that for $n=31$, the result would be the same
as that given by equation~(\ref{eq:origdens}). We make no changes to
the computation of gradients for particles with less than 32
neighbours.

We can now look at the results of using this compensated kernel to
calculate densities of randomly placed gas at different densities as
before -- this is shown as the lower shaded region of
figure~\ref{fig:hydrafix}. Though the scatter is significantly
increased, the mean of densities assigned follows well the true average
densities of the particles down to $\rho/\bar{\rho}_B \sim 0.1$. This
success is slightly tempered by the inevitable dependence of calculated
densities on neighbour distribution when $n$ is small. For example if
the particles are not randomly placed, but distributed evenly on a
grid, then the results of using the compensated kernel to calculate
particles (dashed line in figure~\ref{fig:hydrafix}) is rather
different. The calculated densities are biased lower than their true
densities in a discontinuous fashion since the number $n$ of neighbours
found falls discontinuously from $n=32$ with $h < h_{\rm max}$
(occurring at $\rho/\bar{\rho}_B \gtsima 10$) to 26, 18 and 6 as the
search length $h = h_{\rm max}$, becomes successively smaller in
comparison with the spacing of the particles at decreasing
densities. $n$ finally reaches 0 at $\rho/\bar{\rho}_B = 0$, and thus
evenly spaced particles set up with densities less than this never see
any of their neighbours, and would then according to
equation~\ref{eq:fixdens} be assigned a density simply at the
(compensated) self density level, $\hat\rho_0 = {\cal W}_{ii}/32 =
8/32\pi$.

\begin{figure}
\resizebox{\columnwidth}{!}{\includegraphics{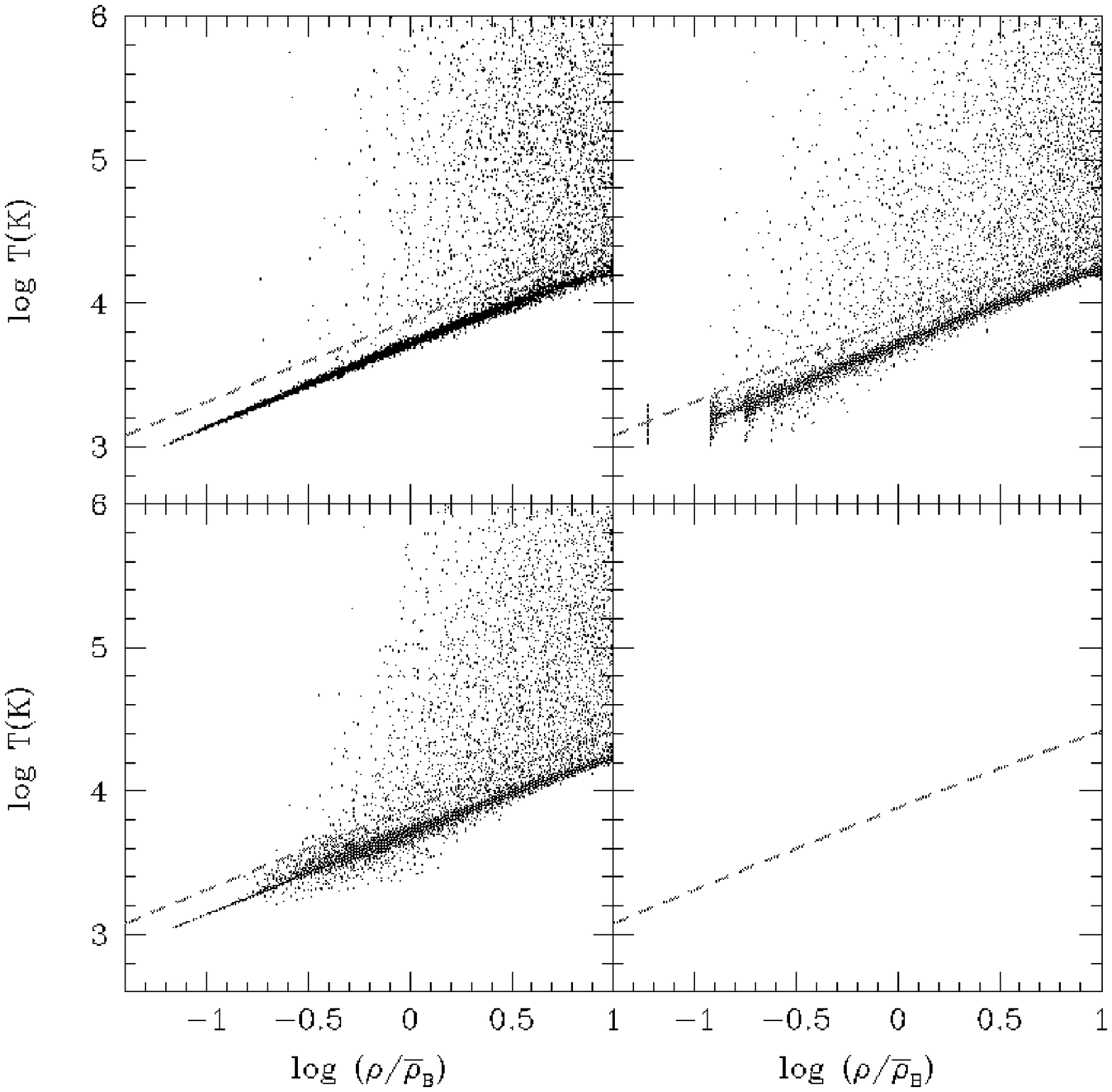}}
\caption{Density-temperature distributions at $z=2$ of runs 22-64-k
using a) \apm, b) \hydra without reconstruction, and c) \hydra after
reconstruction. In panel d) the solid curve shows $\log T_{\rm
min}(\rho)$ for the gas (see figure~\ref{fig:figrt}), and its
approximated power-law form as given by
equation~(\ref{eq:tigmdef}). The dotted line shows the equilibrium
temperature of the gas (where heating balances cooling). The dashed
lines in each panel are identical and the same as those plotted and
defined in figure~\ref{fig:figrt}.}
\label{fig:rtclose}
\end{figure}

In figure~\ref{fig:rtclose}b we show the simulated $\rho-T$
distribution at low densities for one of our runs (H-22-64-k, see
Table~\ref{table:runs} for run labelling) at $z=2$, {\em pre}-density
reconstruction. Since in the simulations the particles are placed
initially on a perturbed grid, many particles do initially have $n$
close to zero, and are assigned densities biased low towards this
compensated $\hat\rho_0$ level, and this situation evidently persists
to z=2, as there is a group of particles fixed all at the same lowest
density. However overall the \hydra $\rho-T$ distribution compares very
well at low densities with the distribution found by \apm
(figure~\ref{fig:rtclose}a) using its exact scheme. This can be
understood since the evolution of gas in the low density IGM with
densities biased low is in fact very similar to the evolution one would
expect with the correct densities. Shock heating processes are not
underestimated as these are negligible for $\rho/\bar{\rho}_B \ltsima
10$ anyway, and pressure effects are also most important for gas
settling into DM halos, where the gas becomes better and better
resolved, and the number of neighbours, and hence the accuracy of the
assigned densities increases.

We also see that the particles in both \apm and pre-reconstruction
\hydra lie on a power-law relation $\log(T) \propto \alpha\log(\rho)$
(whose form is shown in figure~\ref{fig:rtclose}d discussed in detail
in Appendix~\ref{App:IGM}). It is this relation which can be made use
of to post-adjust the particle temperatures at the same time as the
densities in the reconstruction step. Defining pre(post)-reconstruction
densities and temperatures as $\rho_{i(f)}, T_{i(f)}$ we fit the slope
$\alpha$ obeyed by $\rho_i$ and $T_i$, find $\rho_f$ accurately by the
density reconstruction step outlined earlier, and then set $T_f$
according to
\begin{equation}
\log(T_f) = \log T_i + \alpha(\log\rho_f - \log\rho_i)
\end{equation}
\ie proportionally adjusting the temperatures along with the densities
according to the same slope as was obeyed pre-reconstruction. We show
the results of reconstructing the densities and temperatures in this
way in figure~\ref{fig:rtclose}c. It is clear that this procedure
indeed retrieves the low-density distribution so that it compares very
well with the distribution simulated by the exact scheme of \apm, with
only a small induced scatter in temperature evident. (The distribution
of gas below the dashed line was shown earlier in
figure~\ref{fig:rhot}. This distribution also compares well with the
\apm one at low densities.)

\subsection{Calculation of spectra}
\label{sect:spectra}
Given the positions, velocities, densities and temperatures of all SPH
particles at a given redshift, we compute spectra along a given line of
sight through the box as follows. We divide the sight line into $N\sim
1000$ bins of width $\Delta$ in distance $x$ along the sight line. For a
bin $i$ at position $x(i)$ we compute the density and the density
weighted temperature and velocity from:
\begin{eqnarray}
\rho_X(j)   &=& a^3\,\sum_i X(i) {\cal W}_{ij}\\ 
(\rho T)_X(j) &=& a^3\,\sum_i X(i) T(i) {\cal W}_{ij}\\
(\rho v)_X(j) &=& a^3\,\sum_i X(i) (a\dot x(i)+\dot a (x(i)-x(j))
{\cal W}_{ij}\,.\nonumber\\
\end{eqnarray}
where $X(i)$ is the abundance of species $X$ of SPH particle $i$,
assuming ionization equilibrium ($X=\H$, $X=\Hep$; $X=1$ denotes total
gas density, see Appendix~\ref{sect:eqs} for meaning of other symbols).
Labelling bins according to velocity, from zero to $\dot aL$, a bin at
velocity $v(k)$ will suffer absorption from the bin at velocity $v(j)$
by an amount $e^{-\tau(k)}$ where
\begin{eqnarray}
\tau(k) &=& \sigma_\alpha {c\over V_X(j)} \rho_X(j)a\Delta\nonumber\\
&\times & \exp\left(-\left[ {v(k)-v(j)\over V_X(j)}\right]^2\right) /\sqrt{\pi}\label{eq:Gaussian}
\\
V_X^2(j) &=&  {2k_B T_X(j)/m_X}.
\end{eqnarray}
Here, $c$ is the light speed and $V_X$ is the Doppler width of the
species with particles mass $m_X$. The \lya cross section is
$\sigma_\alpha = (3\pi\sigma_T/8)^{1/2} f \lambda_0$, where $\sigma_T=
6.625 \times 10^{-25}$ cm$^2$ is the Thomson cross section, $f=0.41615$
is the oscillator strength and $\lambda_0$ is the rest wavelength of
the transition.  For the hydrogen \lya transition, we take $\lambda_0 =
1215.6{\mbox \AA}$ ($\sigma_\alpha= 4.45\times 10^{-18}$ cm$^2$), for \Hep,
$\lambda_0 = 304.8{\mbox {\AA}}$ ($\sigma_\alpha= 1.12\times 10^{-18}$
cm$^2$. These spectra can converted from \lq velocity\rq~ $v$ to \lq
observed\rq~ wavelength $\lambda$ using
$\lambda=\lambda_0\,(1+z)(1+v/c)$.

\section{Atomic processes and photo-ionization rates}
\label{sect:cooling}
Our simulations include all the physical processes relevant for the
problem of studying primordial gas dynamics in a photo ionised
intergalactic medium, as first collected by Black (1981), and whose
completeness is well addressed elsewhere (see e.g. Katz \etal (1996b)
and references therein). In this Appendix we briefly detail the
functional form of the atomic physics coefficients used. Both \apm and
\hydra use the same set of coefficients, all of which are based on
those collected by Cen (1992) specifically for use in cosmological
hydrodynamic simulations. A few adjustments were made where
improvements in accuracy or economy were found after comparison with
fits used by Efstathiou (1992). Where differences occur these have also
checked out satisfactorily against those quoted using updated atomic
data by Abel \etal (1997), as detailed below.

We calculate the normalised radiative cooling function, ${\cal C}$
(normalised as in equation.~(\ref{eq:thermal}) such that the rate of loss of
thermal energy per unit volume, $\rho du/dt = n^2_{\rm H}{\cal C}$) by
summing the cooling rates whose units and functional dependence on
temperature are given in Table~\ref{table:cooling}, such that
\begin{equation}
{\cal C} = \sum^{11}_{i=1} c_i(T,\rho_B,J(\nu,z),T_{\rm CMB})
\end{equation}
The functions used to fit the $c_i$ are those of Cen (1992), where we
have increased the collisional ionization cooling rates by a factor 2
to offset the reducing effects of the $(1+(T/10^5K)^{1/2})^{-1}$ factor
introduced by Cen to extend the validity of existing fits to higher
temperatures (this factor 2 was analogously applied to our collisional
ionization rates $\Gamma_{e\h,\he,\hep}$ given below).

In order to compute cooling rates as given in Table~\ref{table:cooling}
we need ionic abundances for the different gas species. Normalising
fractional densities to the total hydrogen density, and denoting them
by their standard species nomenclature (\eg $\H = n_{\h}/n_{\rm H}$),
we may write the equations of ionization evolution as
\begin{eqnarray}
{d\H\over dt} &=& \alpha_\hp\, n_e \Hp - \H (\Gamma_{\gamma \h} + 
\Gamma_{e\h}\,n_e) \\
{d\He\over dt} &=& \alpha_\hep n_e \Hep - \He (\Gamma_{\gamma\he} +
\Gamma_{e\he} n_e) \\
{d\Hepp\over dt} &=& \Hep (\Gamma_{\gamma\hep} + \Gamma_{e\hep} n_e) 
-\alpha_\hepp \Hepp n_e\,\nonumber\\
\end{eqnarray}
supplemented with the closing conditions
\begin{eqnarray}
\H+\Hp&=&1\\
\He + \Hep + \Hepp &=& y\\
\Hp+\Hep+2\Hepp &=& e,
\end{eqnarray}
where $y=Y/(m_{He}/m_H(1-Y))$ denotes the helium abundance by number;
$m_{H}, m_{He}$ are the hydrogen and helium atomic mass, $n_e=en_H$ is the
electron number density, and the ionization and recombination rates'
units and functional dependence on temperature are given in
Tables~\ref{table:recomb}, \ref{table:Haardt} \&
\ref{table:photo}. Assuming photo-ionization equilibrium we solve the
resulting set of closed equations iteratively until the fractional
change in all species densities has dropped below 0.001\%.

For the recombination and collisional ionization rates' dependence on
temperature $T$, we used the functional fits listed in
Table~\ref{table:recomb}. Once again these are based on those of Cen
(1992), but in addition to the high temperature factor correction
mentioned above we introduce a factor 0.75 to reconcile the fit quoted
for the \Hp ion recombination rate, with the data points given in
Spitzer (1978). This inconsistency between Cen's fit and Spitzer's data
has already been noted in Rauch \etal (1997, sec. 3.1) where they
introduce a similar factor 0.8, and does not appear to be present for
the \Hepp ion recombination rate coefficient used.

The rate of photo-ionization of any species ion $i$, $\Gamma_{\gamma
i}$, depends on the flux spectrum of ionising ultraviolet background
photons, $J(\nu,z)$ (erg cm$^{-2} $ s$^{-1}$ Hz$^{-1}$ sr$^{-1}$ ), in
the following way:
\begin{eqnarray}
\Gamma_{\gamma i}(z) &=& 
\int^{\infty}_{\nu_i} \frac{J(\nu,z)\sigma_i(\nu)}{h\nu}\,d\nu\,,
\label{eq:gammadef}
\end{eqnarray} 
where $\sigma_i(\nu)$ and $\nu_i$ are the photo-ionization
cross-sections and ionising threshold frequencies respectively for each
species. Similarly, excess energy in electrons ejected through
photo-ionization provides a gas heating mechanism from the UV
background, for which the normalised photo-heating function ${\cal H}$
(again normalised as in equation~(\ref{eq:thermal}) such that the rate
of gain in thermal energy per unit volume, $\rho du/dt = n^2_{\rm
H}{\cal H}$) is given by,
\begin{equation}
{\cal H} = \left(\H\,\epsilon_{\gamma\h}+\He\,\epsilon_{\gamma\he}+
\Hep\,\epsilon_{\gamma\hep}\right)/n_{\rm H}\,.
\label{eq:heat}
\end{equation}
Here the photo-heating coefficients used, $\epsilon_{\gamma}$, are
calculated via an analogous expression to equation~\ref{eq:gammadef}:
\begin{eqnarray}
\epsilon_{\gamma i}(z) = \int^{\infty}_{\nu_i} \frac{J(\nu,z) \sigma_i(\nu)
(h\nu-h\nu_i)}{h\nu}\, d\nu\,,
\label{eq:epsdef}
\end{eqnarray} 
where $h\nu_i$ is the ionization energy.

In this paper we assume two separate models for the UV background
history of the universe. These are implemented by deducing
photo-ionization and photo-heating rates using the above equations. We
describe the actual fits implemented for each coefficient here.

In case of the evolving background UV spectrum computed by Haardt \&
Madau (1996, with deceleration parameter $q_0=0.5$, corresponding QSO
z-evolution, and QSO spectral index equal to 1.5), we performed these
integrations numerically using $\sigma_i(\nu)$ given by Cen (1992). Our
own fits to these integrated photo-ionization and heating rates as a
function of redshift, given in Table~\ref{table:Haardt}, are accurate
at all redshifts to within 8\% and also agree with fits calculated for
the same spectra independently (Haardt, private communication) to
within 2\% (beyond $z=9$ all the rates were set to zero).

As a second more general case we assume a background of ionising
photons with a power-law spectrum of the standard form:
\begin{equation}
\label{eq:powerlaw}
J(\nu) = J_{21} \times 10^{-21}\left({\nu\over\nu_\h}\right)^{-\alpha}
{\rm erg\,s^{-1}\,cm^{-2}\,sr^{-1}\,Hz^{-1}},
\end{equation}
where the background flux is normalised by parameter $J_{21}$ at the \H
Lyman limit frequency $\nu_\h$. We can integrate
equations~\ref{eq:gammadef} \& \ref{eq:epsdef} in general if we
approximate the ion-photon cross-sections, $\sigma_i$, as having simple
power-law dependencies on frequency. This is already the case for
$\sigma_\he$, however for the hydrogenic ions \H and \Hep we use
\begin{eqnarray}
\sigma_i(\nu) &=&
6.3\times10^{-18}{\rm cm}^2 {f_i\over Z^2} \left(\frac{\nu_i}{\nu}\right)^3\,,
\end{eqnarray} 
where $Z$ denotes the atomic number; $f_i\sim 1$ is a dimensionless
constant. By using $f_{\h,\hep} = 1,1.21$ for the photo-ionization
rates and $f_{\h,\hep} = 1.12,1.26$ for the heating rates, the
resulting expressions (given in Table~\ref{table:photo}) agree to 2\%
for a wide range of spectral index, $\alpha$, with numerical
integrations using the exact $\sigma_i(\nu)$.
\begin{table*}
\centering
\begin{minipage}{140mm}
\caption{Cooling rates (ergs cm$^3$\,s$^{-1}$). $z\equiv$ redshift,
$n_H\equiv$ hydrogen number density and $T_n\equiv T/(10^n K)$. $T$ is
in $K$.}
\begin{tabular}{llll}
			 &   & Collisional ionization cooling														&Species          \\
$c_1$    & = & $2.54\times 10^{-21}T^{1/2}e^{-157809.1/T}(1+T_5^{1/2})^{-1}e \H$		   &$\H$             \\
$c_2$    & = & $1.88\times 10^{-21}T^{1/2}e^{-285335.4/T} (1+T_5^{1/2})^{-1}e \He$		&$\He$            \\
$c_3$    & = & $9.90\times 10^{-22}T^{1/2}e^{-631515./T}  (1+T_5^{1/2})^{-1}e \Hep$	   &$\Hep$           \\
			 & 	 & Recombination cooling																	&                 \\
$c_4$    & = & $8.70\times 10^{-27} T^{1/2} T_3^{-0.2} (1+T_6^{0.7})^{-1}e\Hp$			&$\Hp$            \\
$c_5$    & = & $1.55\times 10^{-26} T^{0.3647}e\Hep$											   &$\Hep$           \\
$c_6$    & = & $3.48\times 10^{-26} T^{1/2} T_3^{-0.2} (1+T_6^{0.7})^{-1} e\Hepp$		&$\Hepp$          \\
			 & 	 & Dielectronic recombination	cooling												   &                 \\
$c_7$    & = & $1.24\times 10^{-13} T^{-1.5} e^{-470000./T} (1+0.3 e^{-94000/T})e\Hep$ &$\Hep$           \\
			 & 	 & Collisional excitation cooling														&                 \\
$c_8$    & = & $7.5\times 10^{-19} e^{-118348/T} (1+T_5^{1/2})^{-1}e\H$					   &$\H$             \\
$c_9$    & = & $5.54\times 10^{-17} T^{-0.397} e^{-473638/T} (1+T_5^{1/2})^{-1}e\Hep$	&$\Hep$           \\
			 &   & Bremsstrahlung																			   &                 \\
$c_{10}$ & = & $1.42\times 10^{-27} g_f T^{1/2}e(\Hp+\Hep+4\Hepp)$						 	&$\Hp,\Hep,\Hepp$ \\
$g_f$    & = & $1.1+0.34 e^{-((5.5-\log_{10}(T))^2)/3}$											&                 \\
			 &   & Inverse Compton cooling																   &                 \\
$c_{11}$ & = & $5.406\times 10^{-36}\,(T-2.7(1+z))(1+z)^4 e/n_H$							 	&                 
\label{table:cooling}
\end{tabular}
\end{minipage}
\end{table*}
\begin{table*}
\centering
\begin{minipage}{140mm}
\caption{Recombination and collisional ionization rates in s$^{-1}$,
as a function of temperature $T$(K).}
\begin{tabular}{lll}
						  &   &  Recombination																				 \\
$\alpha_{\hp}$    & = &  $6.30\times 10^{-11}T^{-1/2}T_3^{-0.2}/(1+T_6^{0.7})$							 \\
$\alpha_{\hep}$   & = &  $1.50\times 10^{-10}T^{-0.6353}$														 \\
$\alpha_\hepp$    & = &  $3.36\times 10^{-10}T^{-1/2}T_3^{-0.2}/(1+T_6^{0.7})+\alpha_{\hep}^{(D)}$   \\
						  &	 &	 Dielectronic recombination                                                 \\
$\alpha_{\hep}^{(D)}$&=& $1.9\times 10^{-3}T^{-1.5}e^{-4.7\times10^5/T}(1+0.3e^{-9.4\times 10^4/T})$\\
						  &   &  Collisional ionization																	 \\
$\Gamma_{e\h}$    & = &  $1.17\times 10^{-10}T^{1/2}e^{-157809.1/T}\,(1+T_5^{1/2})^{-1}$ 				 \\
$\Gamma_{e\he}$   & = &  $4.76\times 10^{-11}T^{1/2}e^{-285335.4/T}\,(1+T_5^{1/2})^{-1}$				 \\
$\Gamma_{e\hep}$  & = &  $1.14\times 10^{-11}T^{1/2}e^{-631515/T}\,(1+T_5^{1/2})^{-1}$					
\label{table:recomb}
\end{tabular}
\end{minipage}
\end{table*}

\begin{table*}
\centering
\begin{minipage}{140mm}
\caption{photo-ionization ($\Gamma_\gamma$ in s$^{-1}$) and
photo-heating ($\epsilon_\gamma$ in erg\,s$^{-1}$) rates: Haardt \&
Madau spectrum. ($z\equiv$ redshift)}
\begin{tabular}{llll}
									& \multicolumn{3}{l}{Ionization rates}\\
$\Gamma = e^{x_1+z x_2+z^2 x_3}$  & $x_1$ & $x_2$ & $x_3$\\
$\Gamma_{\gamma\h}$      & -31.04 & 2.795 & -0.5589 \\
$\Gamma_{\gamma\he}$     & -31.08 & 2.822 & -0.5664 \\
$\Gamma_{\gamma\hep}$    & -34.30 & 1.826 & -0.3899 \\
									& \multicolumn{3}{l}{Photo-heating rates}\\
$\epsilon = e^{x_1+z x_2+z^2 x_3}$  & $x_1$ & $x_2$ & $x_3$\\
$\epsilon_{\h}$          & -56.62 & 2.788 & -0.5594\\
$\epsilon_{\gamma\he}$   & -56.06 & 2.800 & -0.5531\\
$\epsilon_{\gamma\hep}$  & -58.67 & 1.888 & -0.3947
\label{table:Haardt}
\end{tabular}
\end{minipage}
\end{table*}

\begin{table*}
\centering
\begin{minipage}{140mm}
\caption{photo-ionization ($\Gamma_\gamma$ in s$^{-1}$) and
photo-heating ($\epsilon_\gamma$ in erg\,s$^{-1}$) rates: power-law
spectra (Compare with equation~(\ref{eq:powerlaw}) for $J_{21}, \alpha$
definition).}
\begin{tabular}{llll}
									& & Ionization rates	      																		\\
$\Gamma_{\gamma\h}$      &=& $1.26\times 10^{-11}J_{21}(3+\alpha)^{-1}$													\\
$\Gamma_{\gamma\he}$     &=& $1.48\times 10^{-11}J_{21}\,0.553^\alpha J_{21}\,										
										 ({1.66\over\alpha+2.05}-{0.66\over \alpha+3.05})$										\\
$\Gamma_{\gamma\hep}$    &=& $3.34\times 10^{-12}J_{21}\,0.249^\alpha\,(3+\alpha)^{-1}$							\\
									& & Photo-heating rates																				\\
$\epsilon_{\h}$          &=& $2.91\times 10^{-22}J_{21} (2 + \alpha)^{-1} (3+ \alpha)^{-1}$						\\
$\epsilon_{\gamma\he}$   &=& $5.84\times 10^{-22}J_{21}\,0.553^\alpha\,													
										 ({1.66\over \alpha+1.05}-{2.32\over\alpha+2.05}+{0.66\over \alpha+3.05})$		\\
$\epsilon_{\gamma\hep}$  &=& $2.92\times 10^{-22}J_{21}\,0.249^\alpha\,(2+\alpha)^{-1}(3+\alpha)^{-1}$		\\
\label{table:photo}
\end{tabular}
\end{minipage}
\end{table*}

\section{Evaluating the IGM temperature}
\label{App:IGM}
In this Appendix we write down the defining equations needed to solve
for the temperature at given density in the high redshift universe
observable through \lya absorption where the net (heating$-$cooling)
time, or `heating time', $t_{\rm heat}$ is equal to the Hubble time
$t_{\rm H}$,
\begin{equation}
t_{\rm heat}(\rho, T, J(\nu), z) = t_{{\rm H}}(z)\,.
\label{eq:tceqth}
\end{equation}
We particularly show that for low densities, this relation approximates
very well a power-law whose amplitude and slope are well-defined, and
dependent only weakly on a few parameters, most notably the (effective)
UV background spectral index (see also Hui \& Gnedin 1997).

We define the Hubble time as follows for an Einstein de-Sitter universe,
\begin{equation}
t_{\rm H} = {1\over(6\pi G\bar{\rho}(z))^{1/2}} = {t'_{\rm
H}\over(1+z)^{3/2}h}\,.\\
\label{eq:hubtime}
\end{equation}
The heating time (or net cooling time) defined in
equation~(\ref{eq:tc}) may be written in general as
\begin{eqnarray}
t_{\rm heat} &=& {3m_{\rm H} k_B \over 2(1-Y)^2\mu\bar{\rho}_0} {T
\over \Omega_B\Delta_B h^2 L(1+z)^3}\nonumber\\ &=& {t'_{{\rm heat}} T
\over \Omega_B h^2\Delta_B L (1+z)^3}\,
\label{eq:cooltime}\,,
\end{eqnarray}
where we use $L = {\cal H}-{\cal C}$ to denote the heating rate (erg
s$^{-1}$ cm$^3$), positive for net heating, and $\Delta_B =
\rho_B/\bar{\rho}_B(z)$. Above and in what follows all fundamental
atomic parameters are collected together into constant factors (each
given a dash), so that the relative contribution from each respective
process together with cosmological parameters may be preserved through
the calculation. It should be noted that although these factors are
denoted by the quantity they attribute to plus a dash, they will not in
general have the same units as this quantity.

Using the equations as laid out in Appendix~\ref{sect:eqs} we may
calculate $L$ for a given density and temperature (and UV background)
and solve equation~(\ref{eq:tceqth}) in a numerical fashion
straightforwardly. We can however simplify the above equation
considerably by approximating the normalised cooling rates in the
low-density regime.

Effectively in the low density region, $\Delta_B < 10$, the heating
rate, $L$, is dominated by \H \& \Hep photo-heating, with a small but
non-negligible Compton cooling contribution. Following the notation of
the previous Appendix we may write the photo-heating component of $L$,
denoted $L_\eps$, simply as
\begin{equation}
L_{\eps} = \H\frac{\eps _{\h}}{n_{\rm H}} + \Hep\frac{\eps
_{\hep}}{n_{\rm H}}\,.
\end{equation}
The \H \Hep and electron fractions above are likely to be extremely
highly photo-ionised and so we may write their abundances accurately as
follows
\begin{equation}
\H = \frac{\alpha_{\hp}n_e}{\Gamma^\gamma_{\h}}\,,\\
\Hep = y\frac{\alpha_{\hepp}n_e}{\Gamma^\gamma_{\hep}}\,,\\
e = 1+2y\,,
\end{equation}
where the recombination coefficients can be well approximated for
temperatures $T< 10^{5}$ K as (see Table~\ref{table:recomb}),
\begin{equation} 
\alpha_{\hp} = \alpha_{\hp}'T^{-0.7}\,, \\
\alpha_{\hepp} = \alpha_{\hepp}'T^{-0.7}\,.
\label{eq:arec}
\end{equation}
Further, if we model the photo-ionization flux $J_{\nu}$ as a power law
according to equation~(\ref{eq:powerlaw}), then the ratios of
photo-heating and photo-ionization coefficients given in
Table~\ref{table:photo} are
\begin{equation} \frac{\eps _{\h}}{\Gamma^\gamma_{\h}} =
\frac{h_P c}{\lambda_{\h}(2+\alpha)}\,,\\ \frac{\eps
_{\hep}}{\Gamma^\gamma_{\hep}} = \frac{h_P c}{\lambda_{\hep}(2+\alpha)}\,,
\end{equation} 
where $\lambda_{\h}$ and $\lambda_{\hep}$ are the ionization
wavelengths for \H and \Hep respectively, and we denote Planck's
constant and the speed of light as $h_P, c$. Drawing these elements
together, the normalised photo-heating rate becomes
\begin{equation}
L_{\eps} = 
\left(\frac{\alpha_{\hp}'}{\lambda_{\h}} + 
y\frac{\alpha_{\hepp}'}{\lambda_{\hep}}\right)
\frac{h_P c(1+2y)T^{-0.7}}{2+\alpha} = 
\frac{L'_{\eps}T^{-0.7}}{2+\alpha}\,.
\label{eq:leps}
\end{equation}
As mentioned previously their is also a small contribution from Compton
cooling, $L_{\rm cc}$, which may be written ($c_{11}$ from
Table~\ref{table:cooling}, $T\gg T_{\rm CMB}$)
\begin{equation}
L_{\rm cc} = 
-\frac{5.406\E{-36}(1+2y)m_h}{(1-Y)\bar{\rho_0}}\frac{T(1+z)}{\Omega_B
h^2\Delta_B} = 
\frac{L'_{\mathrm{cc}}T(1+z)}{\Omega_B h^2\Delta_B}\,.
\label{eq:lcc}
\end{equation}

We are now in a position to solve for the temperature at a given
density where $t_{\rm heat} = t_H$. Substituting for each using
equations~\ref{eq:hubtime}, \ref{eq:cooltime}, \ref{eq:leps} and
\ref{eq:lcc}
\begin{equation}
\frac{t'_{\rm heat} T}{\Omega_B h^2\Delta_B} = 
\left(\frac{L'_{\eps}T^{-0.7}}{2+\alpha} + 
\frac{L'_{\rm cc}T(1+z)}{\Omega_B h^2\Delta_B}\right)
{t'_{\rm H}(1+z)^{3/2}\over h}\,,
\end{equation}
which by multiplying through by $\Omega_B h^2\Delta_B/T$ and
juggling yields the general solution for $T$ at given $\Delta_B$,
\begin{eqnarray}
T &=& T_0\Delta_B^\frac{1}{1.7} \label{eq:tigmdef}\\
T_0 &=& \left[\left(\frac{\Omega_B L'_\eps t'_{\rm H}}{t'_{\rm heat}h}
\frac{(1+z)^{3/2}}{2+\alpha}\right)\bigg/ \left(1-\frac{L'_{\rm
cc}t'_{\rm H}(1+z)^{5/2}}{t'_{\rm heat}h}
\right)\right]^{{1\over1.7}}\,.\label{eq:Tgeneral} 
\nonumber\\
\end{eqnarray} 
The second factor in brackets encloses the contribution of Compton
cooling only. This is found always to be small for intermediate
redshifts. Putting in values for physical constants, using a
cosmological helium fraction, $Y = 0.24$, and
$\mu=(1+4y)/(1+y+e)=0.588$ in the low density limit, we get
\begin{eqnarray}
t'_\h &=& 2.06\E{17}\;{\rm s}\\
t'_{\rm heat} &=& 5.41\E{-11}\;{\rm erg}\;{\rm cm}^3\;{\rm K}^{-1}\\ 
L'_{\eps} &=&1.70\E{-20}\;{\rm erg}\;{\rm s}^{-1}\;{\rm cm}^3\;{\rm K}^{0.7}\\
L'_{\rm cc} &=& -7.31\E{-30}\;{\rm erg}\;{\rm s}^{-1}\;{\rm cm}^3\;{\rm K}^{-1}\,,
\end{eqnarray}
where we have used the following atomic numbers for $L'_{\eps}$:
\begin{eqnarray} 
\lambda_{\h} &=& 911.75 {\mbox {\AA}}\\
\lambda_{\hep} &=& 227.67 {\mbox {\AA}} \\
\alpha_{\hp}' &=& 2.51\E{-10}\;{\rm K}^{0.7}\;{\rm s}^{-1}\\
\alpha_{\hepp}' &=& 1.34\E{-9}\;{\rm K}^{0.7}\;{\rm s}^{-1}\,,
\end{eqnarray}
which may be collected together into equation~(\ref{eq:Tgeneral}) above
to find 
\begin{equation}
T_0 =
\frac{3.92\E{4}(1+z)^{3\over3.4}(\Omega_B h/(2+\alpha))^{1\over1.7}}
{\left(1+{1\over
h}\left(\frac{1+z}{1+9.52}\right)^{5/2}\right)^\frac{1}{1.7}}\;{\rm K}\,,
\label{eq:T0}
\end{equation}
where the main dependencies for $T_0$ are given in the numerator, and the
denominator has the correction due to Compton cooling only.

\end{document}